\documentclass[lettersize,journal]{IEEEtran}
\IEEEoverridecommandlockouts % enable \thanks
\usepackage{etex}

\usepackage[font=footnotesize,caption=false]{subfig} % preload with correct options...

\usepackage{relsize}

\usepackage[dvipsnames,svgnames,usenames]{xcolor}
\usepackage{array}
\usepackage{booktabs,tabularx}
\usepackage{multirow}
\usepackage[binary-units=true,per-mode=symbol,detect-mode=true]{siunitx}
\usepackage{graphicx}
\usepackage[caption=false]{subfig}

\usepackage[T1]{fontenc}
\usepackage{textcomp}
\usepackage[utf8]{inputenc}
\usepackage[final]{microtype}
\usepackage{icomma}
\usepackage{xspace}

\usepackage[tbtags]{amsmath}
\usepackage{amssymb,amsfonts,bm}
\usepackage{mathtools} 
\usepackage{dsfont}
\usepackage{mathrsfs}
\usepackage{accents}
\usepackage{empheq}
\usepackage{nccmath}
\usepackage{balance}

\usepackage{color,colortbl}
\usepackage{calc}
\usepackage{tikz}
\usepackage{pgfplots,pgfplotstable}

\usepackage{pdftexcmds}
\makeatletter
\newcommand{\strequal}[2]{\pdf@strcmp{#1}{#2}==0}
\makeatother

\usepackage{hyperref}
\usepackage[capitalize]{cleveref}
\usepackage{refcount}

\usepackage[inline]{enumitem}
 \usepackage{algorithm}
\usepackage{algpseudocode}
\makeatletter
\newcommand{\algmargin}{\the\ALG@thistlm}
\makeatother
\newlength{\whilewidth}
\algdef{SE}[parWHILE]{parWhile}{EndparWhile}[1]
  {\parbox[t]{\dimexpr\linewidth-\algmargin}{%
     \hangindent\whilewidth\strut\algorithmicwhile\ #1\ \algorithmicdo\strut}}{\algorithmicend\ \algorithmicwhile}%
\algnewcommand{\parState}[1]{\State%
  \parbox[t]{\dimexpr\linewidth-\algmargin}{\strut #1\strut}}

\makeatletter
\newcommand\fs@spaceruled{\def\@fs@cfont{\bfseries}\let\@fs@capt\floatc@ruled
  \def\@fs@pre{\vspace{.05in}\hrule height.8pt depth0pt \kern2pt}%
  \def\@fs@post{\kern2pt\hrule\relax}%
  \def\@fs@mid{\kern2pt\hrule\kern2pt}%
  \let\@fs@iftopcapt\iftrue}
\makeatother

\AtBeginEnvironment{algorithmic}{\footnotesize}
\makeatletter
\newcommand\floatc@ruledn[2]{\small{\@fs@cfont #1} #2\par}
\newcommand\fs@rulednew{\fs@ruled\let\@fs@capt\floatc@ruledn}
 \makeatother
\floatstyle{rulednew}
\restylefloat{algorithm}

\usepackage{glossaries}
\usepackage{ifthen}
\usepackage[noadjust]{cite}
\usepackage{multibib}
\usepackage{bbm}

\usepackage{comment}
\usepackage{todonotes}
\let\legacytodo\todo
\newcommand{\ruggedtodo}[2][]{\tikzexternaldisable\legacytodo[#1]{#2}\tikzexternalenable}
\renewcommand{\todo}[1]{\ruggedtodo[inline]{#1}}
\usepackage{graphicx,color}
\usepackage{textcomp}
\usepackage{xcolor}
\usepackage{hyperref}
\hypersetup{hidelinks=true}

\crefname{equation}{}{}
\crefrangeformat{equation}{~(#3#1#4)--~(#5#2#6)}
\crefmultiformat{equation}{~(#2#1#3)}{ and~~(#2#1#3)}{,(#2#1#3)}{,(#2#1#3)}
\crefrangemultiformat{equation}{#3~(#1)#4--#5~(#2)#6}{, #3~(#1)#4--#5~(#2)#6}{, #3~(#1)#4--#5~(#2)#6}{, #3~(#1)#4--#5~(#2)#6}
\undef\mod
\DeclareMathOperator\mod{mod}

\allowdisplaybreaks[3]

\DeclareSIUnit \dBm {dBm}
\DeclareSIUnit \dBW {dBW}
\DeclareSIUnit \bpcu {bpcu}
\newcommand{\CLASSINPUTtoptextmargin}{2.54cm}
\newcommand{\CLASSINPUTbottomtextmargin}{4.6cm}
\DeclareFontFamily{U}{mathx}{\hyphenchar\font45}
\DeclareFontShape{U}{mathx}{m}{n}{
      <5> <6> <7> <8> <9> <10>
      <10.95> <12> <14.4> <17.28> <20.74> <24.88>
      mathx10
      }{}
\DeclareSymbolFont{mathx}{U}{mathx}{m}{n}
\DeclareMathSymbol{\bigtimes}{1}{mathx}{"91}

\usepackage{amsthm}
\newtheorem{property}{Property}
\newtheorem{assumption}{Assumption}
\pgfplotscreateplotcyclelist{default}{%
	blue,mark=*\\%
	red,mark=star\\%
	teal,mark=square*\\%
	brown!60!black,mark=otimes*\\%
}

\newcolumntype{P}[1]{>{\centering\arraybackslash}p{#1}}

\ifCLASSOPTIONdraftcls
\AtBeginEnvironment{figure}{\def\baselinestretch{1.0}}
\fi
\def\BibTeX{{\rm B\kern-.05em{\sc i\kern-.025em b}\kern-.08em
    T\kern-.1667em\lower.7ex\hbox{E}\kern-.125emX}}

\newacronym{In-F}{In-F}{In-factory}
\newacronym{AI}{AI}{artificial intelligence}
\newacronym{6G}{6G}{sixth generation}
\newacronym{5G}{5G}{fifth generation}
\newacronym{SN}{SN}{sub-network}
\newacronym{HRLLC}{HRLLC}{hyper-reliable, low-latency communication}
\newacronym{URLLC}{URLLC}{ultra-reliable, low-latency communication}
\newacronym{MCS}{MCS}{modulation and coding scheme}
\newacronym{LA}{LA}{link adaptation}
\newacronym{CQI}{CQI}{channel quality indicator}
\newacronym{LSTM}{LSTM}{long short-term memory}
\newacronym{SINR}{SINR}{signal to interference plus noise ratio}
\newacronym{ACK}{ACK}{acknowledgement}
\newacronym{NACK}{NACK}{non-acknowledgement}
\newacronym{BLER}{BLER}{block error rate}
\newacronym{OLLA}{OLLA}{outer-loop link adaptation}
\newacronym{LOS}{LOS}{line-of-sight}
\newacronym{NLOS}{NLOS}{non-line-of-sight}
\newacronym{IIR}{IIR}{infinite impulse response}
\newacronym{eMBB}{eMBB}{enhanced mobile broadband}
\newacronym{TTI}{TTI}{transmission time interval}
\newacronym{IPV}{IPV}{interference power value}
\newacronym{MO}{MO}{modulation order}
\newacronym{ECDF}{ECDF}{empirical cumulative distribution function}
\newacronym{EKF}{EKF}{extended Kalman Filter}
\newacronym{SA}{SA}{sensor-actuator}
\newacronym{EESM}{EESM}{exponential effective signal-to noise-ratio mapping}
\newacronym{ESM}{ESM}{effective SNR mapping}
\newacronym{MIESM}{MIESM}{mutual information effective signal-to noise-ratio mapping}
\newacronym{CSI-RS}{CSI-RS}{channel state information-reference signal}
\newacronym{MSE}{MSE}{mean square error}
\newacronym{IM}{IM}{interference management}
\newacronym{AP}{AP}{access point}
\newacronym{UE}{UE}{user equipment}
\newacronym{CSI}{CSI}{channel state information}
\newacronym{MAC}{MAC}{medium access control}
\newacronym{INR}{INR}{interference-to-noise-ratio}
\newacronym{3GPP}{3GPP}{third generation partnership project}
\newacronym{RDMM}{RDMM}{random directional mobility model}
\newacronym{InF-DL}{InF-DL}{indoor factory with dense clutter and low base station}
\newacronym{SNR}{SNR}{signal to noise ratio}
\newacronym{DSSM}{DSSM}{dynamic state space model}
\newacronym{DL}{DL}{downlink}
\newacronym{UL}{UL}{uplink}
\newacronym{TDD}{TDD}{time division duplexing}
\newacronym{eSNR}{eSNR}{effective signal-to-noise ratio}
\newacronym{RRM}{RRM}{radio resource management}
\newacronym{RAE}{RAE}{relative absolute error}
\newacronym{RSRP}{RSRP}{recieved signal reference power}
\newacronym{CMD}{CMD}{correlation matrix distance}
\newacronym{SD}{SD}{spectral divergence}
\newacronym{WSS}{WSS}{wide sense stationary}
\newacronym{LRS}{LRS}{local region of stationarity}
\newacronym{RS}{RS}{reference signal}
\newacronym{RA}{RA}{resource allocation}
\newacronym{AWGN}{AWGN}{additive white Gaussian noise}
\newacronym{IMF}{IMF}{intrinsic mode functions}
\newacronym{EVT}{EVT}{extreme value theory}
\newacronym{POT}{POT}{peak-over-threshold}
\newacronym{GPD}{GPD}{generalized pareto distribution}
\newacronym{pdf}{pdf}{probability density function}
\newacronym{PI}{PI}{prediction interval}
\newacronym{QR}{QR}{quantile regression}
\newacronym{CQF}{CQF}{conditional quantile function}
\newacronym{CR}{CR}{conformal regression}
\newacronym{CP}{CP}{conformal prediction}
\newacronym{ICP}{ICP}{inductive conformal prediction}
\newacronym{CQR}{CQR}{conformalized quantile regression}
\newacronym{RNN}{RNN}{recurrent neural network}
\newacronym{NN}{NN}{neural network}
\newacronym{ML}{ML}{machine learning}
\newacronym{TX}{TX}{transmission}
\newacronym{TSN}{TSN}{time-sensitive networking}
\newacronym{EtherCAT}{EtherCAT}{ethernet for control automation technology}
\newacronym{ARIMA}{ARIMA}{autoregressive integrated moving average}
\newacronym{NARNN}{NARNN}{nonlinear autoregressive neural network}
\newacronym{LMBP}{LMBP}{Levenberg-Marquardt backpropagation}
\newacronym{STL}{STL}{seasonal and trend decomposition with locally estimated scatterplot smoothing}
\newacronym{GRU}{GRU}{gated recurrent unit}
\newacronym{LS}{LS}{large-scale}
\newacronym{CS}{CS}{conformity score}
\newacronym{FFN}{FFN}{feedforward network}
\newacronym{MLP}{MLP}{multi-layer perceptron}
\newacronym{GEV}{GEV}{generalized extreme value}
\newacronym{FLOPS}{FLOPS}{Floating point operations per second}
\newacronym{OFDM}{OFDM}{orthogonal frequency-division multiplexing
}
\newacronym{iQPTransformer}{iQPTransformer}{\underline{i}nverted \underline{q}uantile \underline{p}atch \underline{t}ransformer}
\newacronym{CPU}{CPU}{central processing unit}
\newacronym{SL}{SL}{split learning}
\newacronym{FP}{FP}{forward propagation}
\newacronym{BP}{BP}{backpropagation}
\author{
    \IEEEauthorblockN{
        Pramesh Gautam\IEEEauthorrefmark{1}~(Graduate Student Member, IEEE),
        Sushmita Sapkota\IEEEauthorrefmark{1},
        Carsten Bockelmann\IEEEauthorrefmark{1}~(Member, IEEE),
        Shashi Raj Pandey\IEEEauthorrefmark{2}~(Member, IEEE),
        and Armin Dekorsy\IEEEauthorrefmark{1}~(Senior Member, IEEE)
    }
    \IEEEauthorblockA{ \\
        \IEEEauthorrefmark{1}Department of Communications Engineering, University of Bremen, 28359 Bremen, Germany \\
        \IEEEauthorrefmark{2}Connectivity Section, Department of Electronic Systems, Aalborg University, Denmark \\
        Emails: \{gautam, bockelmann, dekorsy\}@ant.uni-bremen.de, sapkota@uni-bremen.de, srp@es.aau.dk
    }
    \thanks{
        This work is supported by the German Federal Ministry of Research, Technology and Space (BMFTR) under the grants 16KISK109 (6G-ANNA) and 16KISK016 (Open6GHub).
    }
}

\usepackage[firstpage=true]{background}
\SetBgContents{\textcolor{black}{\footnotesize This work has been submitted to the IEEE for possible publication. Copyright may be transferred without notice, after which this version may no longer be accessible.}}% Set contents
\SetBgPosition{current page.south}% Select location
\SetBgVshift{0.50cm}% Add vertical shift (results in a shift in x direction due to rotation)
\SetBgOpacity{1.0}% Select opacity
\SetBgAngle{0.0}% Select rotation of logo
\SetBgScale{1.0}% Select scale factor of logo
\begin{document}

\title{Extreme Value Theory-based Distributed Interference Prediction for 6G Industrial Sub-networks}

\maketitle  
\begin{abstract}
Interference prediction that accounts for extreme and rare events remains a key challenge for ultra-densely deployed sub-networks (SNs) requiring hyper-reliable low-latency communication (HRLLC), particularly under dynamic mobility, rapidly varying channel statistics, and sporadic traffic. This paper proposes a novel calibrated interference tail prediction framework—a hybrid statistical and machine learning (ML) approach that integrates an inverted quantile patch transformer (iQPTransformer) within extreme value theory (EVT). It captures interference dynamics and tail behavior while quantifying uncertainty to provide statistical coverage guarantees. Its effectiveness is demonstrated by leveraging the estimated interference tail distribution to design predictive, risk-aware resource allocation. In resource-constrained SN scenarios, we introduce the split-iQPTransformer, enabling collaborative training by distributing neural network components between sensor-actuator (SA) pairs and the SN controller, while maintaining minimal performance disparity compared to the centralized iQPTransformer. The framework effectively handles deep fading, random traffic, and time-division duplexing (TDD) misalignments and is resilient to rare and extreme interference events. Extensive evaluations are performed under two mobility models and two realistic SN traffic patterns, using a spatially consistent 3GPP channel model across all scenarios. Experimental results show consistent achievement of block error rate (BLER) targets beyond the 95th percentile in the hyper-reliable regime, significantly outperforming baseline approaches.

\end{abstract}


\section{Introduction}
\label{sec:introduction}
"Network of Networks" has been identified as an essential component of the future \gls{6G} communication infrastructure, offering potential solutions by integrating vertical applications with diverse and heterogeneous requirements \cite{hoffmann2023secure}. To support these requirements, the international telecommunication union (ITU) envisioned the evolution of \gls{5G} towards \gls{6G}, enhancing its capabilities through features such as immersive, massive, and \gls{HRLLC}. These are further complemented by features like ubiquitous connectivity, \gls{AI}-driven communication, and integrated sensing and communication to enable next-generation applications \cite{recommendation2023framework}. One specific category of interest in this work is industrial \glspl{SN}, which are envisioned to support the integration of robots with \gls{HRLLC} requirements in factory environments. These \glspl{SN} comprise \gls{SN} controller that communicate with multiple \gls{SA} pairs, facilitating closed-loop communication between each \gls{SA} pair and its corresponding controller. Such configurations are designed to enable critical industrial functionalities, including motion control, positioning, and proximity sensing—features traditionally provided by wired industrial communication standards such as \gls{EtherCAT}, Profinet, \gls{TSN}. The overarching objective is to achieve comparable levels of reliability, latency, and data throughput, while offering increased flexibility through wireless connectivity \cite{du2022multi,berardinelli2021extreme}.

To meet such extreme requirements, specifically sub-millisecond latency, these industrial \glspl{SN} require high bandwidth for intra-\gls{SN} communication \cite{adeogun2020towards}. In factory environments, the presence of multiple concurrently operating robots results in ultra-dense deployment scenarios. Together, the high bandwidth requirements and ultra-dense deployments cause significant interference, posing a major challenge in environments constrained by limited frequency resources. This issue is particularly pronounced for \glspl{SN}, which must transmit short packets under strict \gls{HRLLC} requirements, further exacerbated by sporadic traffic patterns and the mobility of robots in highly dynamic environments. In such cases, we propose using a predictive interference management technique for dynamic and interference-prone \glspl{SN}, enabling the scheduler to proactively select appropriate resources before interference degrades the performance. Our proposed framework leverages predicted interference estimates that account for uncertainty and extreme channel statistics to ensure reliable and risk-aware communication.

\subsection{Related Works}

Interference prediction has been explored in many existing literature, with some focusing solely on prediction, while others incorporate it into a predictive interference management framework. Schmidt et al. \cite{schmidt2021interference} proposed an \gls{ARIMA}-based Kalman filter for interference prediction, which effectively incorporates channel dynamics and device mobility, assuming Gaussian-distributed covariance. Due to the presence of a finite number of interferers and the impact of dynamic channel statistics, Gaussian assumptions may not be adequate for dynamic industrial \glspl{SN} \cite{prameshjournal,clavier2020experimental}. Furthermore, interference in such environments often exhibits stochastic variation due to sporadic traffic patterns, packet misalignment, and rapidly changing channel conditions such as deep fading. Moreover, \gls{SN} brings additional challenges, since \glspl{SN} have different functionalities compared to terrestrial networks, such as the entire network moving together with their respective \gls{SA} pairs. These factors present considerable challenges for achieving optimal performance when relying on extended Kalman filter frameworks with linear approximations. In \cite{padilla2021nonlinear}, a \gls{NARNN} employing \gls{LMBP} was proposed for interference prediction, effectively exploiting temporal dependencies in interference power values. Following this work, Jayawardhana et al. \cite{jayawardhana2023predictive} and Gunarathne et al. \cite{gunarathne2023decomposition} used empirical mode decomposition to break down the interference power values into \glspl{IMF} and residual series. Moreover, Wei et al. \cite{wei2024joint} employed a hybrid interference prediction technique by first decomposing the interference data using \gls{STL}, and then applying a \gls{GRU} for prediction. These methods require separate machine learning models for each decomposed series, resulting in computationally complex architectures for \glspl{SN}. Additionally, Mahmood et al. \cite{mahmood2020predictive} utilized a discrete-time Markov chain for interference prediction. In that work, a threshold was used to define the area under the interference distribution, thereby enabling the exploitation of tail statistics characterized by the minimum and maximum \gls{INR} under a Gaussian model. Compared to \cite{mahmood2020predictive}, authors of \cite{padilla2021nonlinear} found that  \gls{ML}-based approaches offer more resource-efficient predictions for \gls{RA} prospective. Brighente et al. \cite{brighente2022interference} proposed a kernel-based probability density estimation algorithm and introduced a low-complexity variant for interference prediction in \gls{LA}. Unlike prior approaches that assume Gaussianity, interference in realistic industrial environments often exhibits heavy-tailed statistical properties \cite{clavier2020experimental,prameshjournal}. In this research, we propose a distribution-agnostic, data-driven approach for \textit{learning interference dynamics} using estimates obtained by \gls{SA} pairs based on the \gls{3GPP} channel model. Menholt et al. \cite{Menholt2022Interference} proposed an interference prediction method using the Yule-Walker approach for \glspl{SN}. Marzban et al. \cite{marzban2023deep} proposed that interference prediction is treated as a classification task, aiming to determine the probability that interference power falls within predefined classes. In contrast, we consider interference prediction as a continuous-time regression problem, incorporating rare and extreme interference statistics. In our previous work, we proposed an \gls{LSTM}-based interference prediction technique—with and without federated learning—assuming noise-free estimation and deterministic traffic, as presented in \cite{gautamcooperative} and \cite{pramesh2024int} for In-X \gls{SN} architectures. In \cite{prameshjournal}, we proposed an interference prediction framework to quantify tail statistics using Student-t process regression and an attention-based quantile \gls{LSTM} for a single \gls{SA} pair within a \gls{SN}. However, this approach does not explicitly estimate tail \gls{pdf}, which limits its applicability in designing flexible, risk-aware RA strategies. Interference estimates are often imperfect and the underlying process is stochastic due to the presence of extreme and rare events, leading to potential performance loss. The importance of moving beyond central-limit theorems by including extreme and rare events—utilizing powerful statistical tools such as \gls{EVT}—has been highlighted for \gls{URLLC} in \cite{prameshjournal,bennis2018ultrareliable}. Although not directly focused on interference prediction, \cite{evt_rad_map} estimates rare event statistics by leveraging spatial correlations in radio maps. It integrates non-parametric models with extreme value theory within a Bayesian framework to capture the tail behavior effectively. Salehi et al. utilized a mixture model combining \gls{EVT} and kernel density estimation (KDE) for interference prediction, enabling risk-sensitive \gls{RA} \cite{salehi2025ultra}. In our prior work, we utilized \gls{EVT} to model and predict interference extremes, incorporating the statistical modeling of rare interference events \cite{pramesh_extreme}. Building on this, we now propose a probabilistic framework tailored for \gls{HRLLC}. As we move toward \gls{HRLLC}, it becomes necessary to ensure theoretical statistical guarantees for such tail statistics. To the best of the authors’ knowledge, this is the first work to address statistical coverage guarantees of predicted tail statistics while incorporating interference dynamics for reliable interference prediction in \gls{SN} or \gls{HRLLC} scenarios. Consequently, we propose a probabilistic framework that predicts calibrated tail \glspl{pdf} with a specified level of confidence, enabling risk-aware and proactive resource allocation for interference-prone and resource-constrained \glspl{SN}.

\subsection{Contributions}
% \forest{SRP: the early sentences appear redundant - fits for intro; can we revise this to be precise towards the contributions. }
% Interference prediction that accounts for interference dynamics induced by sporadic traffic, random events, deep fading, and measurement noise is crucial and is often reflected by the heavy-tailed nature of \gls{IPV}.
% Building on this foundation, the present work advances the state-of-the-art by establishing statistical coverage guarantees for predicted tail behavior while explicitly accounting for interference dynamics. 
The key contributions of this work are summarized as follows:

\textit{\textbf{Contribution 1:}} 
We propose a reliable interference threshold prediction framework for dynamically moving \glspl{SN} without making any assumption on the underlying distribution. The predictor requires only valid exchangeability, eliminating dependency on interference distribution assumptions.

\textit{\textbf{Contribution 2:}} 
We introduce an \gls{iQPTransformer} architecture together with \gls{LSTM} to store and retrieve the underlying relationship between interference power values across multi-\gls{SA}-pairs within \glspl{SN}, and to quantify tail statistics for reliable communication. We utilize extreme value theory together with the threshold estimated using the \gls{iQPTransformer} to model the interference tail \gls{pdf}. To ensure statistical guarantees of valid coverage, we further integrate inductive conformal regression, enabling the estimation of a calibrated tail \gls{pdf}—often necessary for designing risk-aware \gls{RA}.

\textit{\textbf{Contribution 3:}} 
For resource-constrained \glspl{SN}, we propose a U-shaped split architecture for the \gls{iQPTransformer}—a distributed interference prediction technique that accounts for their computational limitations. In addition, we predict calibrated tail \gls{pdf} in a fully distributed manner. Results demonstrate that the split model achieves comparable performance with minimal prediction error disparity, while offering a customizable complexity through the distribution of the \gls{iQPTransformer} model based on prior knowledge of computational constraints.

\textit{\textbf{Contribution 4:}} The effectiveness of the proposed predictor is evaluated through resource allocation and metrics related to the prediction error for a spatially consistent \gls{3GPP} channel model with two extreme realistic traffic models compared against baseline methods. In addition, the scalability of the predictor is evaluated by examining its performance across increasing \gls{SA} pairs, thereby assessing its robustness in large-scale deployments.
\subsection{Organization and Notations}

The rest of the paper is structured as follows: the system model is introduced in \cref{chap:systemmodel}. The formulation of the resource allocation problem is shown in \cref{sec.probformulation}. The proposed centralized and distributed probabilistic interference predictors are explained in \cref{sec:ddm} and \cref{sec:distributed_learning}, respectively, with preliminaries in \cref{sec:reliableprediction}. In \cref{sec:simulations}, the simulation setup and results are discussed. The conclusion of this research work is presented in \cref{sec:conclusion}.
 
\textit{Notation: } The bold small and capital letters denote vectors and matrices, respectively. $|.|$ denotes the cardinality of a set; \( \text{Bern}(\cdot) \) and \( \text{Poss}(\cdot) \) denote Bernoulli and Poisson random variables, respectively.

\section{System Model}\label{chap:systemmodel}
\subsection{Sub-network Deployment}
\begin{figure}[t]
\centering
\includegraphics[width=0.50\textwidth]{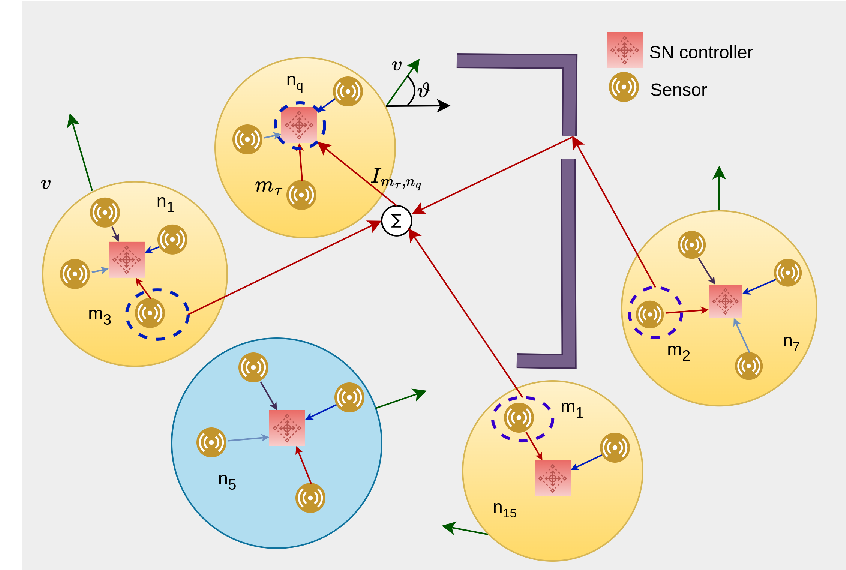}
\caption{Illustration of deployment of $N$  \glspl{SN}. Among the  \glspl{SN} in $N$, there are four  \glspl{SN} from the set $\mathcal{B}=\{ n_{1},n_{q},n_{7},n_{15}\}$ that share the same frequency sub-band. At a specific time slot $\tau$, for example, the $m_{2}$th \gls{SA} of the $n_{q}$th \gls{SN} interferes with the four \gls{SA} pairs ${m_{3}, m_{1}, m_{2}}$ of $\mathcal{B} \setminus {n_{q}}$. The colors of the  \glspl{SN} represent the utilization of distinct sub-bands.}
\vspace{-0.20in}
\label{fig:subnet_layout}
\end{figure}

Consider a set of \glspl{SN}, denoted as $\mathcal{N}=\{n_1,\dotsc,n_N\}$ with a total of $|\mathcal{N}|=N$ \glspl{SN}, located within a factory area $A \subset \mathbb{R}^2$. We assume that these \glspl{SN} move with a constant velocity of \( v \, \text{m/s} \) with a predefined mobility model. We consider two realistic mobility patterns: one in which robots move randomly in different directions—such as moving robotic arms to perform a collaborative tasks within a confined area \cite{adeogun2020towards}—and another in which robots follow predefined trajectories to carry out coordinated operations, such as assembly or inspection, in a factory environment \cite{5G-ACIA}. Each \gls{SN} consists of a set of collocated \gls{SA} pairs to offer closed-loop control communication between these \gls{SA} pairs and the \gls{SN} controller. The sensor sends a sensed data packet with finite blocklength to the \gls{SN} controller over the \gls{UL}. The controller processes this packet and generates a control signal, which is then sent back to the actuator over the \gls{DL} to perform the required action. Each \gls{SN} consists of a set of collocated \gls{SA} pairs $\mathcal{M}_{n_q}= \{m_1,\dotsc,m_M\}~ \text{with} ~|\mathcal{M}_{n_q}|=M,~ \forall ~n_q \in \mathcal{N}$. Those $M$ \gls{SA} pairs are distributed as a binomial point process over a disc of radius $r$, with point density $\frac{1}{\pi r ^2}$. The illustration of deployment is shown in \cref{fig:subnet_layout}.
\subsection{Traffic model within sub-network}\label{traff}
We consider a communication model that facilitates closed-loop control between each \gls{SA} pair and its associated \gls{SN} controller, aiming to replace \gls{EtherCAT} and \gls{TSN} \cite{adeogun2020towards,5G-ACIA}, commonly used in industrial settings—with low-latency and reliable wireless communication \cite{berardinelli2021extreme}. We consider the total available bandwidth \(B\) is partitioned into \(\Omega\) sub-bands, where \(\Omega \ll N\), and a \gls{SN} is assumed to operate in \gls{TDD} mode. It has been assumed that a \gls{SN} is assigned to only a single sub-band at any given \gls{TX} cycle \(t\). However, multiple \glspl{SN} can be allocated to the same sub-band, resulting in an inter-\gls{SN} interference scenario. We adopt a slotted time operation, where each \gls{TX} cycle is indexed by \( t \in \mathbb{N} \) for intra-\gls{SN} communication. The total frame duration, denoted as \(T_{F}\), is divided into two segments: \(T_{\text{UL}}\) and \(T_{\text{DL}}\), corresponding to \gls{UL} communication (from \gls{SA} pair to \gls{SN} controller) and \gls{DL} communication (from \gls{SN} controller to \gls{SA} pair), respectively. Since we focus on the \gls{UL}, the time segment \(T_{\text{UL}}\) is further divided into \(N_{\text{sl}}\), each of duration \(T_{s}\), such that \(T_{\text{UL}} = N_{\text{sl}} \cdot T_{s}\). To comply with \gls{3GPP} specifications, \(T_{s}\) is assumed to be a multiple of an \gls{OFDM} symbol \cite{li2023advanced, cuozzo2022enabling}. Each slot of duration \(T_{s}\) is allocated to a single \gls{SA} pair for \gls{UL} transmission, denoted as \(m_{\tau}\), representing the \(m\)-th \gls{SA} pair active during the \(\tau\)-th slot in the \(t\)-th \gls{TX} cycle \cite{adeogun2020towards}. Intra-\gls{SN} interference is considered minimal due to the use of a scheduler at the \gls{SN} controller, which assigns \gls{SA} pairs in a near-orthogonal manner~\cite{skapoor2018distributedscheduling}. Consequently, it is considered negligible in comparison to inter-\gls{SN} interference, particularly in ultra-dense scenarios~\cite{adeogun2020towards}. Specifically, one \gls{SA} pair $m_{\tau}$ from the set \(\mathcal{M}_{n_{q}}\) scheduled at $\tau$-th slot  interferes with another \gls{SA} pair from each \gls{SN} in the set \(\mathcal{B}\), operating in the same slot and sub-band, where \(\mathcal{B} \subset \mathcal{N}\). In the following, we discuss two distinct traffic models evaluated in this paper.
\subsubsection{Bernoulli isochronous traffic}
The closed-loop control often features a specific type of traffic, referred to as \emph{isochronous}, characterized by the periodic transmission of measurements and control commands \cite{3gpp2019service}. It is assumed that the sensor will send a sensed data packet periodically, which will be processed to generate control data for the actuator in \gls{SN} controller. Assuming the \glspl{SN} are non-synchronized, any \gls{SA} pair $m_{a} \in \mathcal{M}_{n_{q}}$ can be active in time slots within \gls{SN}. A vector $\boldsymbol{M}_{n_{q}} = [m_{1, {n_{q}}}, \dotsc , m_{N_{\text{sl}}, {n_{q}}}]$ represents the active \gls{SA} pairs. Here, each $m_{\tau, {n_{q}}} \in \mathcal{M}_{n_{q}}$ denotes the active \gls{SA} pair of the $n_{q}$th \gls{SN} at time slot $\tau$ during uplink. The set of interfering \gls{SA} pairs is $\mathcal{C}_{m_{\tau}, n_{q}}=\bigcup_{n_j \in \mathcal{B}\setminus \{n_{q}\}}m_{{\tau}, {n_{j}}}$ within the time slot $\tau$. Meanwhile, we incorporate potential randomness in the transmission of packets from the \gls{SA} pair to the \gls{SN} controller. This is modeled as a Bernoulli random variable with a probability of $\chi_{c} \sim \text{Bern}(0, \eta)$ for the scheduled \gls{SA} pair at the slot level, where $\eta$ represents the probability of successful packet transmission when a transmission occurs. 

\subsubsection{Push and Pull-based Traffic}

While the previous model captures periodic, time-triggered communication for closed-loop control, some industrial \glspl{SN} often incorporate both periodic updates and context-aware traffic triggered by real-time events. To capture this behavior more flexibly, we model a combination of push- and pull-based traffic, as formally introduced in \cite{cavallero2024coexistence}. Pull-based traffic is initiated by the \gls{SN} controller to ensure the timely reception of periodic updates—such as sensor readings or status reports. For this purpose, we assign a reserved set of \( N_{\text{sl}}^{\text{ph}} = |\mathcal{M}'_{n_q}| \) slots, where \( \mathcal{M}'_{n_q} \subset \mathcal{M}_{n_q} \) denotes the subset of \gls{SA} pairs that follow deterministic and periodic update transmissions. The remaining \( N - N_{\text{sl}}^{\text{ph}} \) slots are allocated for context-aware random access to accommodate spontaneous events, referred to as push-based traffic. In these slots, any \gls{SA} pair in \( \mathcal{M}_{n_q} \setminus \mathcal{M}'_{n_q} \) can become active based on transmissions initiated by the \gls{SA} pair according to their context, driven by assigned tasks. The activation follows a probability model \( \text{Poss}(0, \lambda) \), where \( \lambda \) represents the traffic intensity associated with random access activity. To incorporate inherent sources of randomness in packet transmission, such as misalignment, we model the occurrence of a packet transmission as a Bernoulli random variable \(\chi_{c} \sim \text{Bern}(\eta)\). This Bernoulli-based model is applied on top of the push-pull traffic at the slot level for a scheduled \gls{SA} pair.
\subsection{Channel and Interference Model}

\begin{figure}
    \centering
    \includegraphics[width=0.95\linewidth]{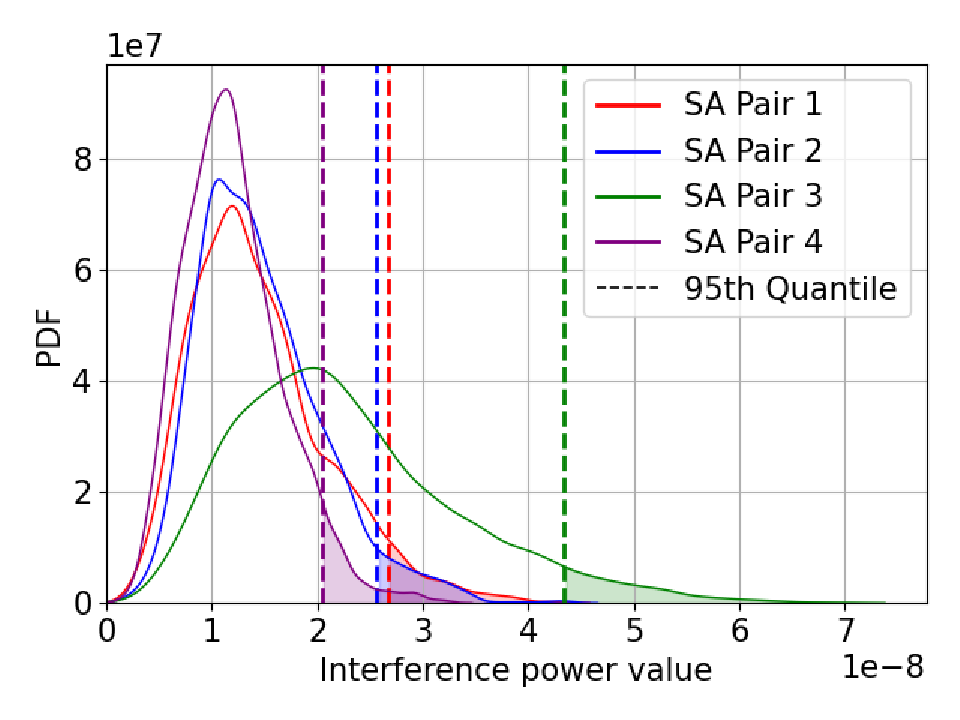}
    \caption{Illustration of \glspl{pdf} of each \gls{SA} pair within the same \gls{SN} under Bernoulli isochronous traffic. The shaded region illustrates the tail \gls{pdf} with a threshold of the 95th quantile.}
    \label{fig:heavy_tail}
\end{figure}

We adopt the channel model from \gls{3GPP} \cite{3gpp2018study}, specifically focusing on the indoor factory scenario characterized by \gls{InF-DL}, to model interference. As the \gls{SN} moves dynamically, the channel conditions can transition between \gls{LOS} and \gls{NLOS} states. These transitions introduce abrupt changes in the channel response due to differences in path loss and \gls{LS} parameters between \gls{LOS} and \gls{NLOS}, resulting in distinct channel realizations. To enable more realistic channel modeling with spatial consistency, we incorporate these abrupt changes using spatially consistent modeling, as recommended in \cite{3gpp2018study}. Further details can be found in \cite{3gpp2018study,prameshjournal}. In this work, we model slot-level interference $\iota_{m_{\tau}, n_{q}}(t)$ based on the spatially consistent \gls{3GPP} channel model as follows \cite{prameshjournal}:
\begin{multline}
\label{eq:interferencemodeling}
\iota_{m_{\tau}, n_{q}}(t) := \sum_{c \in \mathcal{C}_{m_{\tau}, n_{q}}}  P_{c}(t) \cdot \chi_{c}(t) \cdot \biggl(\psi(t) |H_{\text{LOS},c}(t)|^2+ \\ \sqrt{(1- \psi^2(t))} |H_{\text{NLOS},c}(t)|^2\biggr).
\end{multline}
The variable \(\chi_{c}(t)\) represents the \gls{SN} traffic as detailed in previous sub-section, while \(P_{c}(t)\) denotes the transmit power of the interfering \glspl{SN}. The consolidated \gls{LOS} channel, expressed as \(H_{\text{LOS},c}(t) := h_{\text{LOS},c}(t) \cdot l_{\text{LOS},c}(t) \cdot \zeta_{\text{LOS},c}(t)\), consists of three elements: \(h_{\text{LOS},c}(t)\), the small-scale fading coefficient that is correlated across \gls{TX} cycles and follows a Rician 
\gls{pdf} \cite{xiao2006novel}; \(l_{\text{LOS},c}(t)\), the path loss as described in \cite{3gpp2020study}; and \(\zeta_{\text{LOS},c}(t)\), representing the spatially correlated shadowing, which adheres to a log-normal distribution \cite{lu2015effects}. Similarly, the \gls{NLOS} channel, \(H_{\text{NLOS},c}(t)\), is defined with corresponding parameters. The small-scale fading for \gls{NLOS}, denoted \(h_{\text{NLOS},c}(t)\), is correlated across \gls{TX} cycles and follows a Rayleigh pdf. It is combined with the respective path loss, \(l_{\text{NLOS},c}(t)\), and shadowing effect, \(\zeta_{\text{NLOS},c}(t)\), parameterized for \gls{NLOS} conditions. We denote the soft \gls{LOS} state as \(\psi(t) \in [0,1]\), which serves as a time-varying measure to enable smooth transitions between \gls{LOS} and \gls{NLOS} conditions, thereby facilitating a more realistic interference model.  Each \gls{SA} pair estimates interference power during specifically assigned slots within each \gls{TX} cycle, using \gls{CSI-RS} received from the \gls{SN} controller \cite{marzban2024interference,li2023advanced}. The estimated interference is modeled as follows:
\begin{align}
     \tilde{\iota}_{m_{\tau}, n_{q}}(t-1) := \iota_{m_{\tau}, n_{q}}(t-1) + \rho(t-1),
 \end{align}  
where $\rho$ represents estimation noise, assumed to follow a normal distribution $\rho(t-1)\sim \mathcal{N}(0,\sigma^{2})$ with variance $\sigma^{2}$. The inclusion of such estimation error is necessary because obtaining a perfect estimate in a real-time scenario is nearly impossible. These estimated interference power values are stored in the interference power vector, denoted as $\tilde{\boldsymbol{\iota}}_{m_{\tau}, n_{q}}$, for the specific \gls{SA} pair. Overall, the estimated interference power values for for all $\mathcal{M}_{n_q}$ \gls{SA} pairs of $n_q$-th \gls{SN} represented as $\tilde{\mathbf{I}}_{n_{q}} =[\tilde{\boldsymbol{\iota}}_{m_{1}, n_{q}},\dotsc, \tilde{\boldsymbol{\iota}}_{m_{M}, n_{q}}]^{T}$. Interference power often follows non-stationary, correlated, and heavy-tailed \gls{pdf}, as illustrated in \cref{fig:heavy_tail}. This behavior stems from factors like mobility, rapid channel changes (e.g., fading), short packet durations, \gls{TDD}-induced misalignments, estimation errors, and sporadic traffic events \cite{prameshjournal}.
Building on fundamental assumptions observed in realistic \gls{3GPP}, we aim to characterize the statistical properties of interference, with particular emphasis on its tail behavior. As interference often comprises extreme and rare events, it is essential not only to predict tail statistics, but also to establish statistical guarantees based on these predictions for resource allocation that leverages interference knowledge to enable hyper-reliable communication in interference-prone \gls{SN}. Based on this, we formulate a proactive risk-aware resource allocation framework in the following section.

\section{Problem Formulation}
\label{sec.probformulation}
We consider the \gls{RA} problem for uplink communication between the $m_\tau$-th \gls{SA} and the \gls{SN} controller associated with the $n_q^{\text{th}}$ \gls{SN} at a specific \gls{TX} cycle. Our objective is to determine the tight bound for channel usages, $R_{m_\tau, n_q}(t)$, by incorporating the target \gls{BLER} as a reliability constraint. According to finite block length theory \cite{polyanskiy2010channel}, the total number of channel usage that can be transmitted with a decoding error probability of $\varepsilon_{\text{target}}$ for transmission of $D$ information bits in an \gls{AWGN} channel with \gls{SINR} $\hat{\gamma}_{m_\tau, n_q}(t)$ is given as :
\begin{multline}\label{eq:channel_use}
R_{m_\tau, n_q}(t) \approx \frac{D}{\rho(\hat{\gamma}_{m_\tau, n_q}(t))} + \dfrac{{\Theta^{-1}(\varepsilon_{\text{target}})}^2 V(\hat{\gamma}_{m_\tau, n_q}(t))}{2\rho(\hat{\gamma}_{m_\tau, n_q}(t))^2} \cdot \\
\left[ 1 + \sqrt{  1 + \frac{4D \rho(\hat{\gamma}_{m_\tau, n_q}(t))}{\Theta^{-1}(\varepsilon_{\text{target}})^2V(\hat{\gamma}_{m_\tau, n_q}(t))}  } \right],
\end{multline}
where $\rho(\hat{\gamma}_{m_\tau, n_q}) = \log_{2}(1+ \hat{\gamma}_{m_\tau, n_q}(t))$ represents the Shannon capacity of \gls{AWGN} channels under the infinite blocklength regime, $\Theta^{-1}$ is the inverse Q-function, and $V$ is channel dispersion. The selection channel usages $R_{m_\tau, n_q}(t)$ depends solely on $\hat{{\iota}}_{m_\tau, n_q}(t)$, i.e., interference dynamics, as the variation in signal power is insignificant due to the short-range deployment and the fixed positions of the \gls{SN} controller and the \gls{SA} pair. Hence, we decouple interference prediction from the \gls{RA} problem and demonstrate its impact on improving the performance of \gls{RA}. Building on this, we proactively predict the interference for the \( n_q \)-th \gls{SN} at the beginning of the \( t \)-th \gls{TX} cycle, enabling the selection of \( R_{m_\tau, n_q}(t) \) for multiple \gls{SA} pairs. 
To enable risk-aware predictive \gls{RA} in hyper-reliable demanding \gls{SN}, it is crucial to predict interference with rigorous statistical guarantees, particularly in the tail regions of its \gls{pdf}. Interference in such networks often exhibits heavy-tailed behavior and spatio-temporal dependencies between \gls{SA} pairs, making it difficult for conventional predictive methods to capture these statistics. Our goal is to design a framework that estimates interference while accounting for these properties, with a user-defined statistical guarantee level of \(1-\beta\). As a first step, we estimate a probabilistic interference threshold that accounts for local variability and cross-correlations among \gls{SA} pairs within the \gls{SN}. This threshold, predicted at a predefined target quantile \(1-\alpha\), is expressed as
\begin{equation}\label{eq:prob_eq}
    \breve{\mathbf{I}}_{n_q}(t) = f_{1-\alpha}( [\tilde{\mathbf{I}}_{n_q}(t-1), \dotsc, \tilde{\mathbf{I}}_{n_q}(t-S_{w}+1)],\boldsymbol{\theta}),
\end{equation}  
where \( f_{1-\alpha}(\cdot) \) is a probabilistic model that captures the non-linear dynamics of interference by leveraging historical noisy interference observations across multiple \gls{SA} pairs within the same \gls{SN}. The input consists of \(S_w\) correlated past observations, \([\tilde{\mathbf{I}}_{n_q}(t-1), \dotsc, \tilde{\mathbf{I}}_{n_q}(t-S_{w}+1)] \in \mathbb{R}^{S_w \times M}\), and the output is a predicted threshold vector \(\breve{\mathbf{I}}_{n_q}(t) \in \mathbb{R}^{1 \times M}\), where \(M\) denotes the number of \gls{SA} pairs. The parameter \(\boldsymbol{\theta}\) corresponds to the model parameters. Next, we model the tail \gls{pdf} beyond this predicted threshold using \gls{EVT}, which enables characterization of rare and extreme interference events. The estimated tail \gls{pdf} denoted as \(\mathbf{G}(\breve{\mathbf{I}}_{n_q}(t)) := [G_{m_1, n_q}(\breve{\iota}_{m_1, n_q}(t)), \dotsc, G_{m_{N_{\text{sl}}}, n_q}(\breve{\iota}_{m_{N_{\text{sl}}}, n_q}(t))]\), where each \(G_{m_\tau, n_q}(\cdot)\) captures the tail \gls{pdf} for a specific \gls{SA} pair. However, \gls{EVT}-based estimates alone do not provide coverage guarantees, which are crucial for reliable risk-aware \gls{RA}. To address this limitation, we apply \gls{CR} to calibrate the \gls{EVT}-based tail estimates, resulting in a statistically valid tail \gls{pdf} with a theoretical confidence guarantee of \(1 - \beta\), denoted by \(\mathbf{G}^{c}(\breve{\mathbf{I}}_{n_q}(t))\). Finally, we quantify the statistically reliable interference prediction at the desired quantile \(1-\varsigma\) as
\begin{equation}\label{eq:prob_eq_1}
    \hat{\mathbf{I}}_{n_q}(t) = \mathbf{G}^{c}_{1-\varsigma}(\breve{\mathbf{I}}_{n_q}(t)).
\end{equation}
This predicted interference power vector \(\hat{\mathbf{I}}_{n_q}(t)\) used for \gls{RA} (in \cref{eq:channel_use}) to ensure reliable performance. In the next section, we detail the probabilistic interference prediction framework based on this formulation.

\section{Probabilistic Interference Predictor} 
%\forest{Just "Probablisitic Interference Predictor" should work.}}
\label{sec:reliableprediction}
In this section, we first discuss the restructuring of data to ensure exchangeability, followed by preliminaries to facilitate understanding of our proposed probabilistic predictor.

\subsection{Restructuring of Data}
The interference power vector is a non-stationary and correlated process, making reliable prediction particularly challenging. The presence of correlation violates the assumption of exchangeability, which complicates the use of conformal regression for providing statistical guarantees on predictions. However, this correlation is crucial for capturing temporal dynamics. This duality presents a significant challenge: we need to resolve the inconsistency with exchangeability to use conformal regression effectively while preserving the necessary correlation to learn the temporal dependencies of interference power values across successive \gls{TX} cycles. One approach is to utilize the concept of a \textit{stationary interval}, considering that interference can exhibit short-term stationarity, similar to the assumptions in \cite{pramesh2024int,xu2013improving}. Various techniques can be employed to measure the stationary interval of a channel, such as \gls{CMD} \cite{herdin2005correlation}, \gls{LRS} \cite{gehring2001empirical}, and \gls{SD} \cite{georgiou2007distances}. The stationary interval represents the maximum time duration over which the \gls{WSS} assumption holds and can be determined through any method such as  \gls{LRS}, \gls{SD}, or \gls{CMD}. We adopt the stationary interval in terms of interference power rather than the power delay profile of the channel, as commonly used in the literature. Specifically, we use \gls{LRS}, which identifies the largest interval during which the correlation coefficient between two consecutive interference power values exceeds a certain threshold. The stationary interval, \( S_w \), represents the maximum number of \gls{TX} cycles during which the \gls{LRS} correlation remains above a specified threshold \( \phi_c \). It is mathematically defined as:  
\begin{align}\label{eq:lrs}  
    S_{w} = \max \{\Delta \in  \mathbb{N}|\ \wp_{\Delta} \geq \phi_{c}\},  
\end{align}  
where \( \Delta \) signifies the count of correlated samples over successive \gls{TX} cycles, and \( \wp_{\Delta} \) denotes the Pearson correlation coefficient, expressed as \cite{schmidt2021interference}:  
\begin{align}  
    \wp_{\Delta} =  \dfrac{\text{cov}(\tilde{\iota}_{m_\tau, n_q}(t), \tilde{\iota}_{m_\tau, n_q}(t+ \Delta))}{\sqrt{\text{var}(\tilde{\iota}_{m_\tau, n_q}(t)) \cdot \text{var}(\tilde{\iota}_{m_\tau, n_q}(t+\Delta))}}.  
\end{align}  

\begin{figure*}[t]
\begin{minipage}{0.93 \linewidth}
\small
\begin{equation}\label{eq:training_re}
\mathbf{D}_{n_{q}}^{\text{train}}(0) = 
\left(
\begin{bmatrix}
\mathbf{x}_{m_{1},n_q}(0) \\
\mathbf{x}_{m_{2},n_q}(0) \\
\vdots \\
\mathbf{x}_{m_{M},n_q}(0)
\end{bmatrix}, 
\begin{bmatrix}
{y}_{m_{1},n_q}(0) \\
{y}_{m_{2},n_q}(0) \\
\vdots \\
{y}_{m_{M},n_q}(0)
\end{bmatrix}
\right)=
\left(
\begin{bmatrix}
\tilde{\boldsymbol{I}}_{m_{1},n_{q}}(0) & \tilde{\boldsymbol{I}}_{m_{1},n_{q}}(1) & \cdots & \tilde{\boldsymbol{I}}_{m_{1},n_{q}}(S_w-1) \\
\tilde{\boldsymbol{I}}_{m_{2},n_{q}}(0) & \tilde{\boldsymbol{I}}_{m_{2},n_{q}}(1) & \cdots & \tilde{\boldsymbol{I}}_{m_{2}, n_{q}}(S_w-1) \\
\vdots & \vdots & \ddots & \vdots \\
\tilde{\boldsymbol{I}}_{m_{M},n_{q}}(0) & \tilde{\boldsymbol{I}}_{m_{M},n_{q}}(1) & \cdots & \tilde{\boldsymbol{I}}_{m_{M},n_{q}}(S_w-1)
\end{bmatrix}, 
\begin{bmatrix}
\tilde{\boldsymbol{I}}_{m_{1},n_{q}}(S_w) \\
\tilde{\boldsymbol{I}}_{m_{2},n_{q}}(S_w) \\
\vdots \\
\tilde{\boldsymbol{I}}_{m_{m},n_{q}}(S_w)
\end{bmatrix}
\right)
\end{equation}
\begin{equation}\label{eq:test_re}
\mathbf{D}_{n_q}^{\text{test}}(j) = 
\left(
\begin{bmatrix}
\mathbf{x}_{m_{1},n_q}^{*}(j) \\
\mathbf{x}_{m_{2},n_q}^{*}(j) \\
\vdots \\
\mathbf{x}_{m_{M},n_q}^{*}(j)
\end{bmatrix}, 
\begin{bmatrix}
{y}_{m_{1},n_q}^{*}(j) \\
{y}_{m_{2},n_q}^{*}(j) \\
\vdots \\
{y}_{m_{M},n_q}^{*}(j)
\end{bmatrix}
\right)
=
\left(
\begin{bmatrix}
\tilde{\boldsymbol{I}}_{m_{1},n_{q}}(j) & \tilde{\boldsymbol{I}}_{m_{1},n_{q}}(j+1) & \cdots & \tilde{\boldsymbol{I}}_{m_{1},n_{q}}(j+S_w-1) \\
\tilde{\boldsymbol{I}}_{m_{2},n_{q}}(j) & \tilde{\boldsymbol{I}}_{m_{2},n_{q}}(j+1) & \cdots & \tilde{\boldsymbol{I}}_{m_{2},n_{q}}(j+S_w-1) \\
\vdots & \vdots & \ddots & \vdots \\
\tilde{\boldsymbol{I}}_{m_{1},n_{q}}(j) & \tilde{\boldsymbol{I}}_{m_{1},n_{q}}(j+1) & \cdots & \tilde{\boldsymbol{I}}_{m_{1},n_{q}}(j+S_w-1)
\end{bmatrix}, 
\begin{bmatrix}
\tilde{\boldsymbol{I}}_{m_{1},n_{q}}(S_w+j) \\
\tilde{\boldsymbol{I}}_{m_{1},n_{q}}(S_w+j) \\
\vdots \\
\tilde{\boldsymbol{I}}_{m_{1},n_{q}}(S_w+j)
\end{bmatrix}
\right)
\end{equation}
\normalsize
\end{minipage}
\end{figure*} 
Here, \( \text{cov}(\cdot) \) denotes covariance, while \( \text{var}(\cdot) \) represents variance. Within \( S_w \), the interference power values across consecutive \gls{TX} cycles are correlated, leading to non-exchangeable data. To ensure adherence to the principle of exchangeability between instances, these correlated samples are separated into different instances. Consequently, a sliding window-based approach, with the length of the window \( S_w \) calculated from \cref{eq:lrs}, is employed to restructure the interference estimates, forming a training dataset of \( L_{\text{train}} \) instances, denoted by \( \mathbf{D}_{\text{train}}:= (\mathbf{X}, \mathbf{y}) \). Here, the training instances \( \mathbf{X}:= \left[\mathbf{X}(0), \ldots, \mathbf{X}(L_{\text{train}}-1) \right]^{\top} \in \mathbb{R}^{L_{\text{train}} \times S_w \times M} \) represent the input data, with each instance containing \( S_w \) historical interference estimates, such as $[\iota_{m_\tau, n_q} (t)| i = 1, \dotsc S_{w}]$ for $M$ \gls{SA} pairs. We denote each instance within a stationary interval based on \cref{eq:lrs} as \( \mathbf{X}(t) \in \mathbb{R}^{S_w \times M} \). The corresponding labels for each instance \(\mathbf{y}(t) \in \mathbb{R}^{1 \times M}\), associated with each \gls{SA} pair are given by \(\mathbf{Y} = \left[ \mathbf{y}(0), \ldots, \mathbf{y}(L_{\text{train}} -1) \right]^{\top} \in \mathbb{R}^{L_{\text{train}} \times M}\), as illustrated for a single instance in \cref{eq:training_re}. After restructuring the training data \( \mathbf{D}_{\text{train}} \), it is further divided into two mutually exclusive subsets: the training set \( \mathbf{D}_{\text{train}} \) and the calibration data \( \mathbf{D}_{\text{cal}}:= (\mathbf{X}^{\text{c}}, \mathbf{y}^{\text{c}}) \) with $L_{\text{cal}}$ calibration instances, satisfying \( \mathbf{D}_{\text{traincal}} = \mathbf{D}_{\text{train}} \cup \mathbf{D}_{\text{cal}} \) and \( \mathbf{D}_{\text{train}} \cap \mathbf{D}_{\text{cal}} = \emptyset \). For brevity reasons, we note that $\mathbf{X}^{\text{c}} = \left[\mathbf{X}^{\text{c}}(j), \ldots, \mathbf{X}^{\text{c}}(j+L_{\text{c}}-1) \right]^{\top} \in \mathbb{R}^{L_{\text{c}} \times S_w \times M} $ represents the calibration input data, and \( \mathbf{Y}^{\text{c}} = \left[\mathbf{y}^{\text{c}}(j), \ldots, \mathbf{y}^{\text{c}}(j+L_{\text{c}}-1) \right]^{\top} \in \mathbb{R}^{L_{\text{c}} \times M} \) denotes the corresponding calibration labels. After training, the model is evaluated on a new set of previously unseen observations, denoted as \( \mathbf{D}_{\text{test}} := (\mathbf{X}^{*}, \mathbf{Y}^{*}) \), where \( \mathbf{X}^{*} = \left[\mathbf{X}^{*}(j), \ldots, \mathbf{X}^{*}(j+L_{\text{test}}-1) \right]^{\top} \in \mathbb{R}^{L_{\text{test}} \times S_w \times M} \) represents the test input data, and \( \mathbf{Y}^{*} = \left[\mathbf{y}^{*}(j), \ldots, \mathbf{y}^{*}(j+L_{\text{test}}-1) \right]^{\top} \in \mathbb{R}^{L_{\text{test}} \times M} \) denotes the corresponding test labels, illustrated for one instance in \cref{eq:test_re}.

\subsection{Preliminaries in Probabilistic Interference Prediction}
Instead of predicting a single point estimate, this work focuses on estimating a calibrated predictive tail \gls{pdf}, thereby enabling a more comprehensive characterization of uncertainty, particularly in the extremes. To facilitate a clear understanding of the proposed methodology, we briefly review \gls{QR}, \gls{CR}, and other related terminology that will be used throughout the proposed technique.
The predicted interference at a specific quantile, denoted by \(\breve{\iota}_{m_{\tau}, n_{q}}\), is later utilized to characterize the tail statistics of the interference distribution. In this context, coverage is defined as the probability that the true interference power does not exceed the predicted quantile. The prediction is considered statistically valid if the coverage probability for a new test point \((\mathbf{X}^{*}, \mathbf{y}^{*}) \in \mathbf{D}_{\text{test}}\) for each \gls{SA} pair is guaranteed to be at least \(1 - \beta\)-th quantile:
\begin{align}\label{eq:valid_cov}
    \mathbb{P}\left[\mathbf{y}^{*}(t) \preceq \hat{\mathbf{y}}(t)\right] \geq 1 - \beta.
\end{align}
% \forest{how is $\beta$ defined here.} 
Here, \(\mathbf{y}^{*}(t)\) denotes the interference power vector at time \(t\), and \(\hat{\mathbf{y}}(t)\) represents the corresponding predicted interference power vector. The inequality \(\mathbf{y}^{*}(t) \preceq \hat{\mathbf{y}}(t)\) is interpreted elementwise, indicating that each predicted value is greater than or equal to its label for each \gls{SA} pair. To predict interference at a specified quantile from data, we utilize \gls{QR} as a first step due to its ability to model conditional distributions and capture asymmetric uncertainty. In the following, we briefly review \gls{QR} and its formulation in the context of interference prediction.

\subsubsection{Quantile Regression}
The estimated confidence interval of \gls{QR} with empirical \gls{CQF} is expressed as follows:
\begin{align}\label{eq:qr_interval}
    \hat{O}_{\alpha}(\mathbf{X}(t)) = [f_{{\alpha}}(\mathbf{X}(t)),f_{{1-\alpha}}(\mathbf{X}(t))].
\end{align}
As observed in \cref{eq:qr_interval}, the confidence interval is derived from a finite sample subspace, and ensures capturing the local variability of the input space to approximate the underlying function $f_{\alpha}(\cdot)$ with the $\alpha^{th}$ quantile of the predictor. In this work, we focus on predicting the upper quantile $f_{1-\alpha}(\mathbf{X}(t))$, which can be learned by minimizing the asymmetric pinball loss. The pinball loss of an observation pair $(\mathbf{X}(t),\mathbf{y}(t))$ is defined as \cite{wang2019probabilistic}:
\begin{small}
\begin{align}\label{eq:pinball}
\mathcal{L}_{1-\alpha,t} =  
\begin{cases}
  \alpha(f_{1-\alpha}(\mathbf{X}(t))-\mathbf{y}(t)), & f_{1-\alpha}(\mathbf{X}(t))\geq \mathbf{y}(t),\\
  (1-\alpha)(\mathbf{y}(t) -f_{1-\alpha}(\mathbf{X}(t))), & f_{1-\alpha}(\mathbf{X}(t)) < \mathbf{y}(t),
\end{cases}
\end{align}
\end{small}

where $f_{1-\alpha}(\mathbf{X}(t))$ denotes the $1-\alpha$-th quantile, % of the predictor \forest{bring this defn of alpha right next after (11)}
and $\mathbf{y}(t)$ corresponds to the response of the $t$-th \gls{TX} cycle. The pinball loss can be used as an objective function for training deep neural networks to predict a specific quantile defined with $1-\alpha$. Such  \gls{QR} models the interference dynamics by accounting for the tail statistics of predicted interference power values. However, it is important to note that the actual coverage may not guarantee to assure the specified confidence level $1-\beta$  with \cref{eq:qr_interval} \cite{romano2019conformalized}.  As a consequence, when applied in the context of use cases that demand extremely high reliability, such as \gls{SN}, the lack of statistical theoretical guarantees renders only \gls{QR} inadequate. This necessitates the exploration of solutions that can rigorously address these shortcomings.

\subsubsection{Conformal Regression}
Conformal Regression~(\gls{CR}), often referred to as \gls{CP}, is a prominent non-parametric learning framework for uncertainty quantification that offers reliable prediction intervals with coverage guarantees for finite samples \cite{vovk2005algorithmic, xu2023conformal}. Unlike Bayesian frameworks, it does not rely on distributional assumptions, making it a model-agnostic method that can be seamlessly integrated with any prediction algorithm to provide uncertainty of predictions \cite{stankeviciute2021conformal}. It was formally introduced in \cite{shafer2008tutorial}. It has been shown that under the assumption of exchangeability in data, this method generates valid marginal coverage for various applications \cite{xu2023conformal}. Many \gls{CP} methods have been developed to quantify uncertainty in predictive models. These approaches can be classified as either transductive or \gls{ICP}. The transductive inference method is often utilized, as different data realizations are iteratively presented to the learning algorithm. However, this approach is not only computationally expensive but also less practical, particularly for interference prediction, where multiple realizations of interference estimates may not be easily accessible \cite{kath2021conformal}. As a result, it is unsuitable for \gls{SN} scenarios. On the other hand, \gls{ICP} addresses this limitation, although it necessitates the division of training data into two disjoint sets \cite{papadopoulos2002inductive}.  The training set $\mathbf{D}_{\text{train}}$ is used to train the underlying model, while the calibration set $\mathbf{D}_{\text{cal}}$ is employed for computing nonconformity scores based on \cref{algo:conformal_regression}. In Algorithm~\ref{algo:conformal_regression}, line 8, we estimate the $1-\beta$-th quantile for each \gls{SA} pair, which may differ across pairs, and we exploit this in a distributed manner. 
\begin{algorithm}
\caption{ Conformal Prediction for multivariate interference prediction}
\label{algo:conformal_regression}
\begin{algorithmic}[1]
\State \textbf{Input:} A trained model \( f_{1-\alpha}(\cdot, \theta) \) for the \(1 - \alpha\) quantile, an exchangeable calibration dataset \(\mathbf{D}_{\text{cal}}\), and a calibration threshold \(1 - \beta\)
\State \textbf{Output:} Conformity score $\textbf{CS}_{n_{q}}$
\State Initialize $\mathcal{R}_{n_{q}} = \{\}$
\For{$i= 1,2,\dotsc, L_{\text{c}}$}
\State $\hat{\mathbf{y}}^{c}(t) \leftarrow f_{1-\alpha}(\mathbf{X}^{\text{c}}(t), \theta)$
\State $\mathcal{R}_{n_{q}} \leftarrow \mathcal{R}_{n_{q}} \cup [|\mathbf{y}^{c}(t)-\hat{\mathbf{y}}^{c}(t)|]$
\EndFor
\State $\textbf{CS}_{n_{q}} \leftarrow z_{1-\beta}(\mathcal{R}_{n_{q}})$ \Comment{$\textbf{CS}_{n_{q}} \in  \mathbb{R}^ {1 \times M}$; $1-\beta$ quantile of $\mathcal{R}_{n_{q}}$}
\State return $\textbf{CS}_{n_{q}}$ // For all M-SA pairs
\end{algorithmic}
\end{algorithm}

Furthermore, to ensure the validity property outlined in property~\ref{pp:Validity}, it is essential for the data to adhere to the exchangeability based on assumption~\ref{as:Exchang}. This requirement becomes particularly critical when dealing with the interference power vector, which represents correlated non-stationary data. Maintaining data exchangeability is crucial for preserving essential underlying relationships, which, in turn, guarantees marginal coverage. This is made possible through our proposed restructuring of the training and calibration datasets, which satisfies the following property and assumption (dropping subscript $n_{q}$ for clarity): 

\begin{property}{}\label{pp:Validity}
\textbf{(Validity)} Under the exchangeability assumption, any conformal predictor will return the prediction region $z_{\beta}(\mathbf{X}(t))$ such that the probability of error $\mathbf{y}(t+1) \notin z_{\beta}(\mathbf{X}(t+1))$ is not greater than $\beta$. This can be mathematically written as:
\begin{align}
    \mathbb{P}[\mathbf{y}(t+1) \prec z_{1-\beta}(\mathbf{X}(t+1))|\mathbf{D}_{\text{train}}] \succeq 1-\beta
\end{align}
\end{property}

\begin{assumption}{}\label{as:Exchang}
\textbf{(Exchangeability)} In a dataset of $l$ observations $\mathbf{D}_{\text{train}} = \{(\mathbf{X}(t),\mathbf{y}(t))\}_{t=1}^{l}$, any of its $l!$ permutations are equiprobable.
\end{assumption}

The \gls{ICP} provides the conditional \gls{PI} for predicted quantiles from \gls{CR} $\hat{\mu}(\mathbf{X})$ as
\begin{multline}\label{eq:conf_cp}
    C_{\beta}(\mathbf{X}(t)) =  [\hat{\mu}(\mathbf{X}(t))- z_{1-\beta}(\mathcal{R},~\mathbf{D}_{\text{cal}}),\\ \hat{\mu}(\mathbf{X}(t)) +  z_{1-\beta}(\mathcal{R},~\mathbf{D}_{\text{cal}})],
\end{multline}
where $\hat{\mu}(\mathbf{X}(t))$ is the prediction made by underlying predictive model, $\mathcal{R}$ is the set of residual computed prediction of samples $\mathbf{D}_{\text{cal}}$, and \gls{CR} $z_{1-\beta}(\mathcal{R},~\mathbf{D}_{\text{cal}})$ is $1-\beta$ quantile of residual. Moreover, such a prediction is model-agnostic, meaning that any model can be used to predict $\hat{\mu}(\mathbf{X}(t))$; one example is $\hat{\mu}(\cdot):= f_{1-\alpha}(\cdot)$. It is clear from \cref{eq:conf_cp} that \gls{CR} yields a \gls{PI} with fixed width, which does not adapt to interference dynamics \cite{jensen2022ensemble}. Although the fixed-width \gls{PI} has known limitations, we leverage its advantage of providing tail \gls{pdf} with guaranteed marginal coverage within our probabilistic prediction framework. This enables us to predict a calibrated tail \gls{pdf} that explicitly incorporates interference dynamics, while still maintaining statistical coverage guarantees through \gls{CR}.

\begin{figure*}
    \centering
    \includegraphics[width=0.75\linewidth]{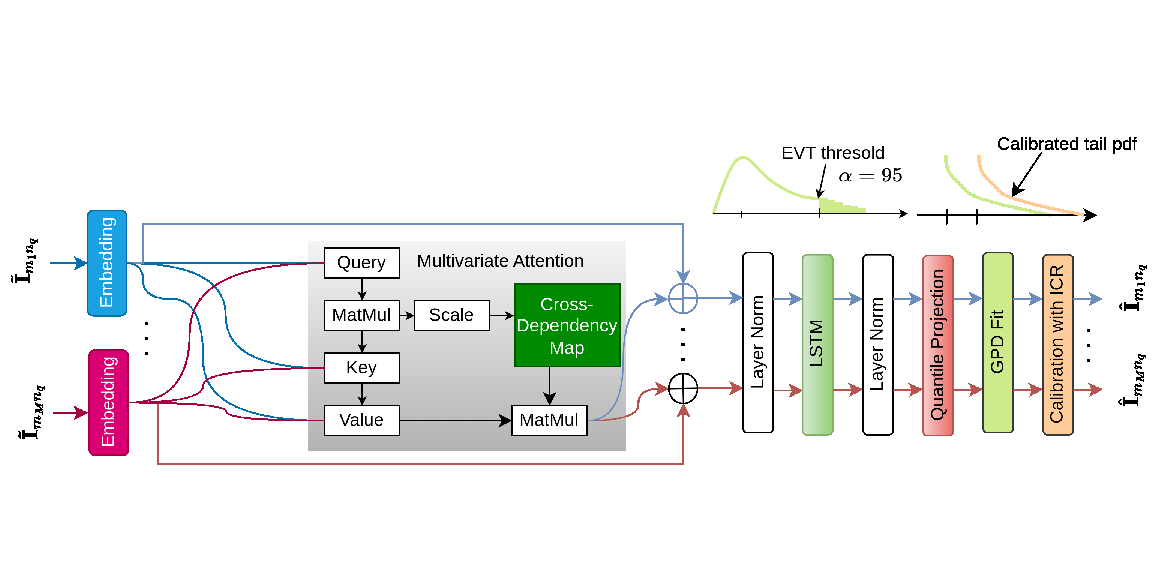}
    \caption{Proposed multi-\gls{SA} pair interference predictor with \gls{EVT}-based \gls{iQPTransformer} for  tail \gls{pdf}  estimation.}
    \label{fig:inverted_data}
\end{figure*}
\section{Centralized Probabilistic Interference Predictor}\label{sec:ddm}
In this section, we introduce a novel framework for calibrated tail \gls{pdf} prediction. First, we predict interference thresholds using the centralized technique illustrated in \cref{fig:inverted_data}. This is followed by tail \gls{pdf} estimation using \gls{EVT}, and finally, a calibrated tail \gls{pdf} is predicted to ensure reliable and uncertainty-aware inference.
\subsection{Inverted Quantile Transformer based Predictor}
Conventional interference prediction techniques often lack the flexibility to accurately model interference dynamics due to fixed statistical assumptions. For example, Gaussian-based models exhibit rapidly decaying tails and therefore fail to account for the probability of rare but impactful interference events. Similarly, reliance on second-order statistics proves insufficient in the presence of sporadic traffic patterns and rare or extreme channel conditions \cite{prameshjournal}. Consequently, \gls{ML}-based approaches are gaining prominence as more flexible and data-driven alternatives, particularly in scenarios where closed-form analytical models are either intractable or lack generalizability across heterogeneous deployment settings. In this work, we propose and adopt an inverted Transformer architecture to predict interference thresholds at defined quantiles, specifically tail statistics, by leveraging spatio-temporal relationships with the following features:
\begin{itemize} 
\item capturing temporal dynamics based on historical estimates for each \gls{SA} pair and enrich the model by exploiting dependencies across other \gls{SA} pairs, 
\item designing an innovative token that aligns with the exchangeability property of \gls{CR},
\item quantifying tail statistics (e.g., the 95th percentile) to incorporate the inherent impact of sporadic events and estimation errors in interference prediction.
\end{itemize}

\begin{figure}
    \centering
    \includegraphics[width=0.99\linewidth]{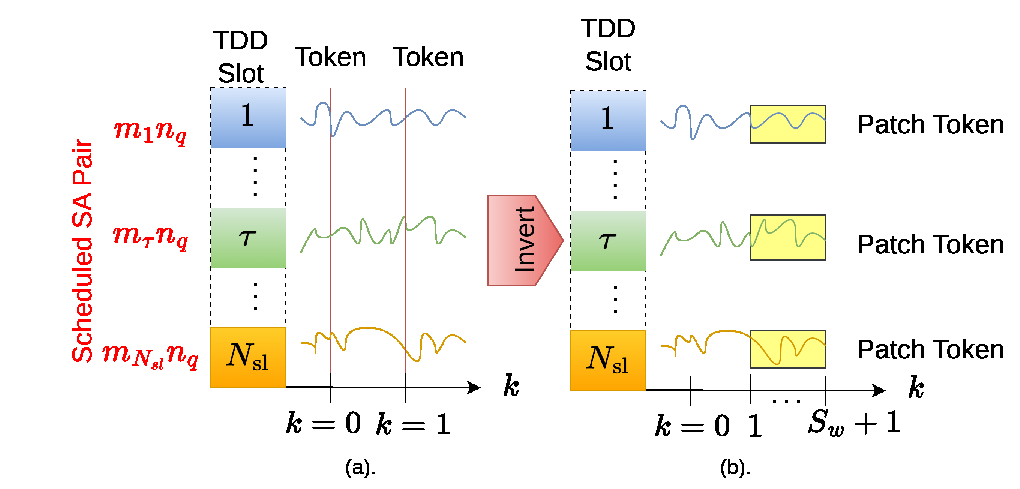}
    \caption{a). Token representation of vanilla transformer. b). Inverted token representation of interference power values for $M$ \gls{SA} pairs for training.}
    \label{fig:inverted_token}
\end{figure}

The transformer network can be employed in prediction tasks to capture global dependencies across temporal tokens. However, its overall performance heavily relies on the design of these tokens, which must effectively encode relevant temporal and spatial information. The spatial distribution of \gls{SA} pairs introduces unique interference dynamics, which often result in \gls{SA} pair-specific statistics as shown in \cref{fig:heavy_tail} and in \cite{pramesh2024int}. Moreover, predicting interference in the presence of multiple \gls{SA} pairs as a multivariate inference necessarily requires a meaningful attention map, as it may result in unreliable predictions, which are unsuitable for use cases such as \gls{SN}, which demand \gls{HRLLC}. One promising approach is the inverted transformer (iTransformer), proposed by \cite{liu2023itransformer}, which effectively learns time-series data by leveraging a meaningful attention mechanism within its inverted structure using temporal tokens. However, directly applying this technique violates the principle of exchangeability, which is essential for its later use in \gls{CR}. To address this, we propose a token design inspired by the patch token approach in \cite{nie2022time}, combined with the inverted token structure. This redesigned token preserves exchangeability. The standard transformer token based on vanilla transformer \cite{vaswani2017attention} and our proposed patch-inverted token of length $S_{w}$, highlighted with yellow rectangular boxes for multiple \gls{SA} pairs, are illustrated in \cref{fig:inverted_token} (a) and (b), respectively. Such inverted token design helps to exploit the temporal dynamics of each \gls{SA} pair individually while enriching the representation by effectively exploiting spatial dependencies across other \gls{SA} pairs. While the iTransformer attends to relevant features in the data, it inherits limitations from traditional \glspl{FFN} used in Transformer architectures, particularly their inability to effectively capture temporal dependencies essential for sequential token processing. To mitigate this, we integrate an \gls{LSTM} to store and retrieve temporal relationships. To further account for uncertainty in interference prediction, we incorporate \gls{QR}. Building on these enhancements, we propose the \gls{iQPTransformer}—a Transformer-based architecture designed to adaptively predict interference thresholds across \gls{TX} cycles, capturing variability while preserving the impact of uncertainty. The \gls{iQPTransformer} leverages attention mechanisms to capture cross-correlations among \gls{SA} pairs and integrates a \gls{LSTM} network to model the non-linear temporal dynamics of each \gls{SA} pair within the same \gls{SN}. This integration enables a more robust and comprehensive inference of multivariate temporal interference dynamics. The architecture comprises four key components—an embedding layer, a multivariate attention layer, layer normalization, and an \gls{LSTM} layer—which work collectively to enhance predictive performance in interference-aware scheduling and resource allocation tasks shown in \cref{fig:inverted_data}.

\subsubsection{Embedding Layer}
It processes the noisy interference estimate \( \tilde{\mathbf{X}}(t-1) \) and maps it into a lower-dimensional latent space for compact and expressive representation. Unlike standard transformers, where temporal tokens are embedded per time step or \gls{TX} cycle, the inverted transformer adopts an inverted embedding strategy, where embeddings are generated across the temporal window for each \gls{SA} pair. This is formulated for each \gls{SA} pair with slot level interference power values as  
\begin{equation}
    \Upsilon_{m_\tau, n_q} = \text{Embedding}({\mathbf{x}}_{m_\tau,n_q}(t-1)),
\end{equation}  
where the embedding is implemented via a \gls{MLP}, enabling non-linear transformation to extract key temporal features while reducing redundancy and noise. The embedded token \( \Upsilon_{m_\tau, n_q} \in \mathbb{R}^D \) captures interference characteristics of the full lookback window for each \gls{SA} pair. The resulting set of \( M \) embedded tokens, denoted as \( \mathbf{H} = \{\Upsilon_{m_1, n_q}, \dotsc, \Upsilon_{m_M, n_q}\} \), serves as the input to the attention layer, allowing the model to learn spatial correlations across \gls{SA} pairs efficiently.

\subsubsection{Multivariate attention layer}
The multivariate attention layer is designed to capture dependencies among the interference dynamics of multiple \gls{SA} pairs. This allows the model to account for user-centric interference driven by the spatial distribution of \gls{SA} pairs, as well as cross-correlation effects stemming from their relative positioning and impact of spatially correlated shadowing.  Utilizing the comprehensively extracted latent space from the embedding layer, the attention layer computes attention scores \( a_{ij} \) for each \gls{SA} pair based on the query \( \mathbf{Q} \), key \( \mathbf{K} \), and value \( \mathbf{V} \) matrices, which are obtained by linear projections of the embedded tokens \( \mathbf{H} \in \mathbb{R}^{M \times D} \), such that  
\begin{equation}
\mathbf{Q} = \mathbf{H} \mathbf{W}_Q,\quad \mathbf{K} = \mathbf{H} \mathbf{W}_K,\quad \mathbf{V} = \mathbf{H} \mathbf{W}_V,
\end{equation} 
where \( \mathbf{W}_Q, \mathbf{W}_K, \mathbf{W}_V \in \mathbb{R}^{D \times d_k} \) are learnable projection matrices. The attention score is computed as  
\begin{equation}
    a_{ij} = \text{softmax}\left( \frac{\mathbf{Q}_i \mathbf{K}_j^\top}{\sqrt{d_k}} \right),
\end{equation}  
where \( d_k \) is the dimensionality of the key vectors. Each token is normalized along its feature dimension, ensuring that the entries reflect the relationships among the individual \gls{SA} pairs' dynamics. The output of the attention layer is then given as  
\begin{equation}
    \mathbf{A}_{\text{attn}} = \sum_j a_{ij} \mathbf{V}_j.
\end{equation}  
% \forest{what is $\alpha_{ij}$ here and where is A?}
The score map \( \mathbf{A}_{\text{attn}} \in \mathbb{R}^{M \times M} \) captures these pairwise relationships, effectively modeling the multivariate dependencies among the tokens. Consequently, variates with stronger correlations are assigned higher weights during the next step in the interaction with the value matrix \( \mathbf{V} \) \cite{liu2023itransformer}. This mechanism enables the model to focus on the relevant contributions of different \gls{SA} pairs, ensuring an accurate representation across multiple \gls{SA} pairs while enhancing interpretability and revealing multivariate dependencies \cite{liu2023itransformer}.
\subsubsection{Layer Normalization}

Layer normalization has been utilized to stabilize and improve the training of \gls{iQPTransformer}. It adjusts the activations in each layer by normalizing the mean and variance of each token \cite{liu2023itransformer}, formulated as  
\begin{equation}
     \text{LN}(\boldsymbol{H}) = \Bigg\{\frac{H_{m_\tau, n_q} - \mu_{H_{m_\tau, n_q}}}{\nu_{H_{m_\tau ,n_q}}} \,\Big|\, \tau = 1, \dotsc, N_{\text{sl}}\Bigg\},
\end{equation}  
where \( \mu_{H_{m_\tau, n_q}} \) and \( \nu_{H_{m_\tau, n_q}} \) denote the mean and standard deviation computed over the feature dimension of each embedded token \( H_{m_\tau, n_q} \). This operation has been shown to be effective for handling non-stationary time series by maintaining a stable activation scale across layers \cite{kim2021reversible}. It ensures that the model processes each \gls{SA} pair’s representation consistently, contributing to robustness and compactness during training.

\subsubsection{LSTM Layer}

\Gls{RNN} is a popular technique for interference prediction due to its ability to capture long-term dependencies in interference data \cite{padilla2021nonlinear,pramesh2024int,marzban2024interference,jayawardhana2023predictive,wei2024joint}. They have been tested and proven effective in various scenarios, showcasing their capability to model complex temporal relationships in the data. \gls{LSTM} networks are a specialized type of \gls{RNN} designed to capture long-term dependencies in sequential data. They address challenges like the vanishing and exploding gradient problems by employing a gated mechanism \cite{Manaswi2018}. It consists of cells that retain or discard information, regulated by input, forget, and output gates. These gates determine which information should be preserved or ignored over time, enhancing the model's ability to learn complex temporal relationships \cite{sak2014long}. In the context of interference prediction, the \textit{\gls{LSTM} layer} plays a pivotal role in capturing long-term dependencies in the interference data \cite{pramesh2024int}. At each time step \( t \), the hidden state is updated as \( \Psi_t = o_t \odot \tanh(\mathbf{c}_t) \), where \( o_t \) is the output gate, \( \mathbf{c}_t \) is the cell state, and \( \odot \) denotes element-wise multiplication. This mechanism allows \glspl{LSTM} to maintain and utilize relevant information from previous time intervals, significantly improving prediction accuracy. A detailed explanation of \gls{LSTM} can be found in \cite{sak2014long}.

\subsubsection{Quantile Projection}
Interference power values often exhibit uncertainty due to channel statistics, traffic patterns, mobility, and impairments such as \gls{TDD} misalignment. Accurately quantifying and accounting for such uncertainty is crucial for reliable communication. Therefore, we aim to predict interference at a $1-\alpha$ quantile, such as the 95th quantile, beyond the mean for reliable communication, based on heuristic evaluations from \cite{prameshjournal}. Thus, \textit{quantile projection} is performed by utilizing the \textit{pinball loss function} \cref{eq:pinball} to estimate dynamic probabilistic thresholds for interference prediction without assuming the distribution of the target interference power values.

The overall computational complexity of the \gls{iQPTransformer} is \( \mathcal{O}(S_w \cdot M^2 \cdot d_e + S_w \cdot M \cdot H^2) \), where \( S_w \) represents the sequence length, \( M \) is the number of \gls{SA} pairs, \( d_e \) is the embedding dimension, and \( H \) is the \gls{LSTM} hidden size. This complexity is primarily driven by the quadratic dependence on \( M \) due to the attention mechanism and the linear dependence on \( S_w \) from sequential processing in the \gls{LSTM}.
Utilizing this network, interference can be predicted at arbitrary quantiles. In the following subsection, we use this probabilistically predicted threshold for tail estimation using \gls{EVT}.

\begin{algorithm}
\caption{Calibrated interference tail estimation and predictive interference management}
\label{algo:reliable_dist_free}
\begin{algorithmic}[1]
\State Collect interference power vector for all \gls{SA} pair as $[\tilde{\mathbf{I}}_{ m_{1} n_k},\dotsc,\tilde{\mathbf{I}}_{m_{M} n_q }] $ 
\State Restructure and split data into $\mathbf{D}_{\text{train}},~\mathbf{D}_{\text{cal}},~\mathbf{D}_{\text{test}}$
\Statex
\State   \textbf{Input:}  Target \gls{BLER} $\varepsilon_{\text{target}}$, Signal Power $\mathbf{S}$, $\mathbf{D}_{\text{train}},~\mathbf{D}_{\text{cal}},~\mathbf{D}_{\text{test}}$
\State \textbf{Input:} Model $f(\cdot)$, target quantile $1-\alpha$
\State \textbf{Output:} Channel Usage $\mathbf{R}$ and achieved \gls{BLER} $\boldsymbol{\varepsilon}$
\Statex
\For{each epoch $i$ from $1$ to $E$}
\If {Resource Constrained Mode == Yes}
\State Train \gls{iQPTransformer} with different $1-\alpha$
\State Train \textbf{head model} and \gls{FP} to \gls{SN} controller
\State Train \textbf{body model} and \gls{FP} to \gls{SA} pair 
\State Train \textbf{tail model} and calculate train-loss 
\State Backpropagate from Client Tail to Server to Client Head.
\State Update $\theta_{p}^{n_{q}}$
\Else
\State Train \gls{iQPTransformer} with different quantile $1-\alpha$
\State Update $\theta_{p}^{n_{q}}$
\EndIf
\EndFor
\State
\State Exceedances $\boldsymbol{\psi}$ from training data and fit GPD $\mathbf{G}(\boldsymbol{\psi})$ from \cref{eq:gpd_fit}
\State Calculate the uncertainty interval by Inductive Conformal Prediction
\State $\mathbf{\text{CS}}:=\mathbf{z}_{1-\beta}(\mathcal{R}, 
    ~\mathbf{D}_{\text{cal}}) = [{z}_{1,1-\beta},\dotsc , {z}_{M,1-\beta}]$ with CP
\Statex  $\text{CPInterval}(f_{\theta_{p}^{n_{q}}}, \mathbf{D}_{\text{cal}}, \beta)$ // Algorithm 1
\For{$t = 1, \dotsc , L_{\text{test}}$}
 \State \textbf{Interference Prediction Phase:}
 \State $\breve{\mathbf{I}}_{n_q}(t) =$ $f_{1-\alpha}(\mathbf{X}^{*}(t))$ 
 \State  $\mathbf{G}_{n_q}(\mathbf{\tilde{I}}_{n_q}(t)) := \breve{\mathbf{I}}_{n_q}(t) + \mathbf{G}(\boldsymbol{\psi})$  \Comment{Estimate tail pdf using \cref{eq:gpd_fit}}
 \State Estimate calibrated interference  tail pdf: $\mathbf{G}^{c}(\breve{\mathbf{I}}_{n_q}(t))$ 
 \State \textbf{Resource Allocation Phase:}
  \State Predict interference $\mathbf{\hat{I}}_{n_q}(t) \leftarrow \mathbf{G}_{1-\varsigma}^{c}(\breve{\mathbf{I}}_{n_q}(t))$
  \State $\mathbf{R}_{n_q}(t): = [R_{m_1, n_q}(t), \dotsc , R_{m_{M}, n_q}(t)]$ based on \cref{eq:channel_use}
  \State  Calculate Achievable \gls{BLER} \cite{mahmood2020predictive}
  \State  Update $\mathbf{R}\leftarrow \mathbf{R}_{n_q}(t)$  and $\boldsymbol{\varepsilon} \leftarrow \boldsymbol{\varepsilon}_{n_q}$
\EndFor
\end{algorithmic}
\end{algorithm}

\subsection{Calibrated interference tail pdf  estimation with EVT}
Although our proposed \gls{iQPTransformer} is capable of capturing tail statistics—by enabling the prediction of arbitrary quantiles—we incorporate calibrated \gls{EVT} for the following reasons:
\begin{itemize} 
\item to assess the stochastic properties of interference from extreme events, such as deep fading and rare traffic, by modeling the tail \gls{pdf}, 
\item to capture rare deviations not accounted for by standard \gls{ML}/\gls{iQPTransformer}-based predictors, 
\item to provide statistical guarantees for rare events, supporting the design of a risk-aware \gls{RA} for hyper-reliability demanding \gls{SN}.
\end{itemize}
\gls{EVT} based approaches are typically built around deriving block maxima using the Fisher–Tippett–Gnedenko theorem or extracting peak values above (or below) a certain threshold from continuous interference estimates based on the Pickands–Balkema–De Haan theorem \cite{haan2006extreme,bennis2018ultrareliable}. Both theorems asymptotically characterize the statistics of extreme events, providing a principled approach for analyzing ultrareliable communication, such as failures with extremely low probabilities \cite{bennis2018ultrareliable}. These extreme events must be considered before introducing risk-aware predictive interference management with ultrareliable or even stringent requirements, such as those that demand \gls{HRLLC}. \gls{EVT} is used to model the distribution of extreme events, employing the \gls{GPD} or \gls{GEV} distribution to capture the tail statistics that arise from rare, stochastic interference behaviors. Further details on \gls{EVT} can be found in \cite{haan2006extreme,bennis2018ultrareliable}. We utilize the \gls{POT} mechanism, which models exceedances of interference over a dynamically defined threshold, offering a more fine-grained and data-efficient approach for capturing rare but impactful deviations crucial for \gls{HRLLC} scenarios. To achieve this, we apply \gls{EVT} through the \emph{\gls{POT}} method, where exceedances over the predicted $\alpha$th quantile threshold \( \breve{\iota}_{m_\tau, n_q}(t) \) are modeled using the \emph{Generalized Pareto Distribution~(\gls{GPD})}, characterized by the shape parameter \( \xi \) and scale parameter \( \sigma \), with the cumulative distribution function as follows \cite{haan2006extreme}:
\begin{align} \label{eq:gpd_fit}
G_{m_\tau, n_q}(\breve{\iota}_{m_\tau, n_q}(t)) = 1 - \left( 1 + \xi \frac{\tilde{\iota}_{m_\tau, n_q}(t) - \breve{\iota}_{m_\tau, n_q}(t)}{\sigma} \right)^{-\frac{1}{\xi}},
\end{align}
where \( \sigma > 0 \), exceedances \( \tilde{\iota}_{m_\tau, n_q}(t) > \breve{\iota}_{m_\tau, n_q}(t) \), and \mbox{\( 1 + \xi \frac{\tilde{\iota}_{m_\tau, n_q}(t) - \breve{\iota}_{m_\tau, n_q}(t)}{\sigma} > 0 \)}. The parameter \( \xi \) plays a crucial role in determining the tail behavior of the distribution, with \( \xi > 0 \) indicating a heavy-tailed Pareto-type distribution, \( \xi = 0 \) corresponding to a light-tailed exponential distribution, and \( \xi < 0 \) representing a bounded beta-type distribution. Moreover, the scale parameter \( \sigma \) is dependent on the threshold \( \breve{\iota}_{m_\tau, n_q}(t) \), such that changes to the threshold result in adjustments to \( \sigma \) according to the relationship:

\begin{equation}
    \sigma(\breve{\iota}_{m_\tau, n_q}^*(t)) = \sigma(\breve{\iota}_{m_\tau, n_q}(t)) + \xi(\breve{\iota}_{m_\tau, n_q}^*(t) - \breve{\iota}_{m_\tau, n_q}(t)),
\end{equation}
where \( \breve{\iota}_{m_\tau, n_q}^*(t) > \breve{\iota}_{m_\tau, n_q}(t) \), this ensures that the model can accommodate threshold variations while maintaining consistency in the exceedance behavior. As referenced in line 24 of \cref{algo:reliable_dist_free}, we apply maximum likelihood estimation to determine the optimal values of the shape parameter $\xi$ and the scale parameter $\sigma$ for each \gls{SA} pair. These parameters are then used to estimate the interference tail \gls{pdf} via \cref{eq:gpd_fit}, using a threshold predicted by the \gls{iQPTransformer}, which incorporates local interference dynamics. Although this \gls{pdf} captures extreme and rare interference statistics, it lacks theoretical statistical guarantees. To address this, we leverage a restructuring technique—outlined in \cref{sec:reliableprediction}—that ensures exchangeability between instances (assumption~\ref{as:Exchang}) and thereby satisfies the property of validity (property~\ref{pp:Validity}). Building on this, we further integrate the estimated tail \gls{pdf} into the model-agnostic framework using \gls{CR}, which enables us to provide statistical guarantees. To predict a calibrated tail \gls{pdf}, we use the calibration dataset $\mathbf{D}_{\text{cal}}$ to compute residuals $\boldsymbol{\varepsilon}_{m}$ for each \gls{SA} pair. These residuals are used to calculate conformity scores based on a $(1 - \beta)$ threshold, which are then stored in $\mathbf{CS}$, as shown in lines 22–23 of \cref{algo:reliable_dist_free}. This calibration step quantifies uncertainty without assuming any specific distribution. Finally, the conformity scores are used to refine the tail \gls{pdf} via conformal regression, as formalized in \cref{eq:conf_cp} as follows:
\begin{multline} \label{eq:cal_pdf}
    G^{c}_{m_\tau, n_q}(\breve{\iota}_{m_\tau, n_q}(t)) = [G_{m_\tau, n_q}(\breve{\iota}_{m_\tau, n_q}(t))-  z_{1-\beta}(\mathcal{R}, 
    ~\mathbf{D}_{\text{cal}}), \\ G_{m_\tau, n_q}(\breve{\iota}_{m_\tau, n_q}(t))+z_{1-\beta}(\mathcal{R},~\mathbf{D}_{\text{cal}})]
\end{multline}
The estimated calibrated interference tail \gls{pdf} is  stored in $\mathbf{G}^{c}(\breve{\mathbf{I}}_{n_q}(t)) = [G^{c}_{m_\tau, n_q}(\breve{\iota}_{m_1 n_q}(t)), \dotsc, G^{c}_{m_{N_{\text{sl}}} n_q}(\breve{\iota}_{m_\tau, n_q}(t))]$. The $1-\varsigma$-quantile of the calibrated interference tail \gls{pdf} is employed for \gls{RA}, effectively accounting for rare and extreme interference events overall summarized in \cref{algo:reliable_dist_free}. One significant flexibility of our proposed framework is that it allows selective modeling of any desired quantile of the tail distribution, enabling tailored, risk-aware interference prediction aligned with reliability constraints. Additionally, when \( \nu = 1 \), the predicted interference does not incorporate \gls{EVT} and instead yields a calibrated interference threshold without requiring prior distributional assumptions. The details corresponding to lines 7–15 will be discussed in the next subsection. In the following, we extend the centralized framework to a distributed setting suitable for resource-constrained \glspl{SN} by restructuring the inverted \gls{iQPTransformer} with a Split-\gls{iQPTransformer}, enabling decentralized prediction of the calibrated interference tail while preserving the probabilistic structure of the predictor.

\section{Distributed Probabilistic Interference Predictor} \label{sec:distributed_learning}
\begin{figure}[!hbt]
    \centering
    \includegraphics[width=0.99\linewidth]{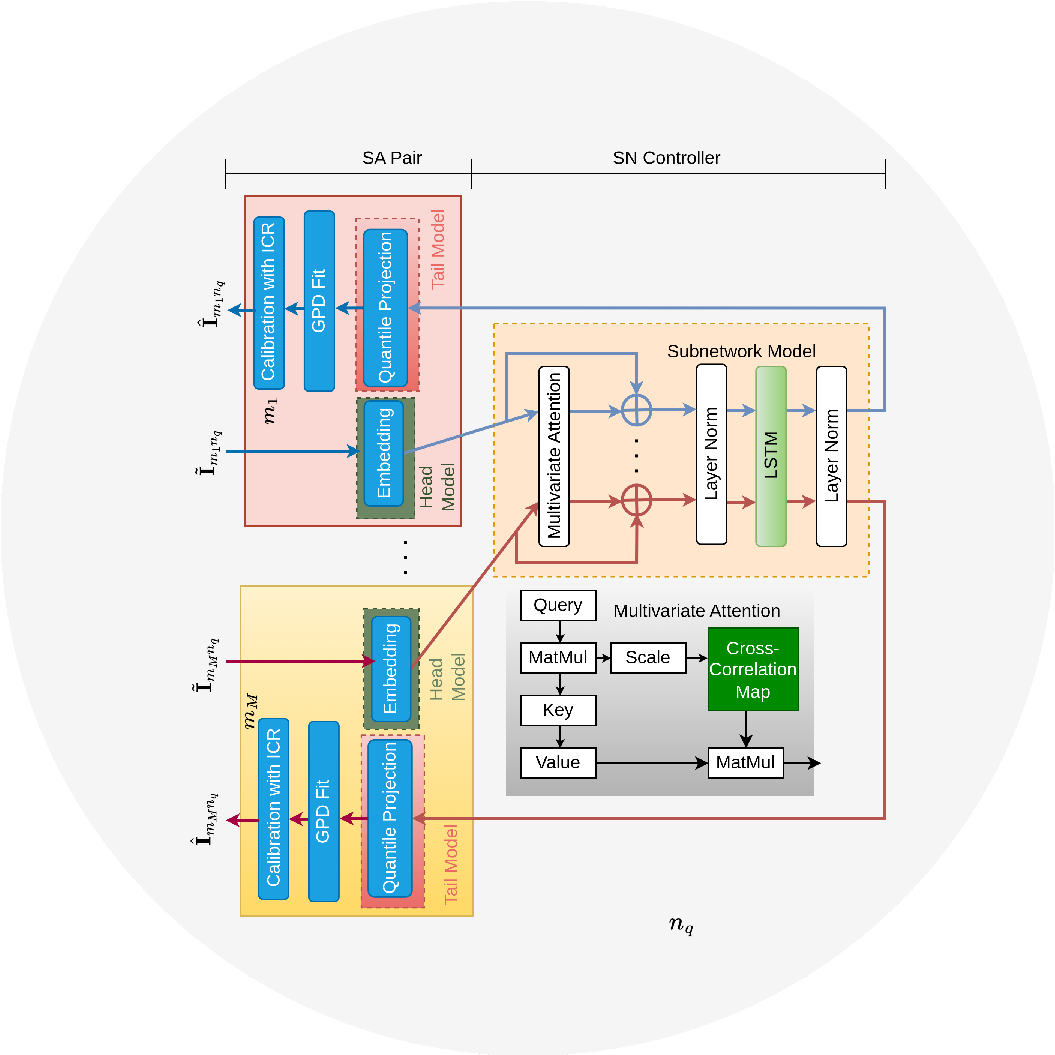}
    \caption{Proposed Split-\gls{iQPTransformer} Architecture}
    \label{fig:lstm_split}
\end{figure}

Interference prediction is inherently a distributed inference problem, as it requires estimating interference directly at spatially distributed \gls{SA} pairs. Leveraging this characteristic, we propose a distributed interference prediction technique tailored for \glspl{SN} architectures utilizing raw interference power values, particularly for use cases involving computationally constrained \glspl{SN}. To utilize advanced models like the \gls{iQPTransformer} for scalable and accurate interference prediction, it may be necessary to redesign the system architecture with careful consideration of computational complexity. We propose a framework that supports this goal by offering computational flexibility through a distributed processing paradigm. Specifically, embedding and initial feature extraction are performed locally at the \gls{SA} pairs, while the resulting activations are sent to the \gls{SN} controller. This allows the controller to focus on resource-intensive tasks, such as modeling temporal dynamics and capturing cross-dependencies between multiple \gls{SA} pairs, enabling efficient deployment of complex interference predictors without compromising performance. The \gls{SL} architecture consists of two parts: a client and a server. In the most basic configuration of split learning, each client trains a portion of a deep neural network up to a specific layer, referred to as the "cut layer". The outputs produced at this cut layer, also called \textit{smashed data}, are then transmitted to another entity, which can be a server. The server takes the smashed data as input and propagates forward until the server-side network's last layer. Subsequently, after calculating the loss, the server starts \gls{BP}, computes the gradients up to the cut layer, and sends the smashed data's gradient back to the client. The client then completes the remaining steps of the \gls{BP} process within its own network. This iterative process continues until the distributed split learning network converges. Since we are interested in predicting interference at the \gls{SA} pair, we need to extend the vanilla split approach to a U-shaped split architecture for uplink. In this paper, we propose a U-shaped Split-\gls{iQPTransformer} configuration that eliminates the need to share either interference data or the corresponding labels during the training and inference phases. This configuration is shown in \cref{fig:lstm_split}. The proposed U-shaped Split-\gls{iQPTransformer} consists of one server model in the \gls{SN} controller and one head and tail model for each \gls{SA} pair. On the client side, we assume that each \gls{SA} pair has the computational capabilities to train the head and tail models, as well as the ability to perform \gls{FP} and \gls{BP}. Similarly, the \gls{SN} controller executes the body model, computes the model parameters, and facilitates the transfer of activations and \gls{BP} as shown in \cref{fig:lstm_split}.

The functionalities of the head, tail, and server model of the proposed U-shaped Split-\gls{iQPTransformer} are as follows:

\subsubsection{\gls{iQPTransformer} Split Client~(Head Model)}
During \gls{FP}, the head model—which includes an embedding layer—transforms raw interference values into lower-dimensional latent representations. This embedding captures essential temporal features while reducing data dimensionality, ensuring that only relevant information is transmitted to the \gls{SN} controller. In the \gls{BP} phase, once the gradients corresponding to the cut layer are received from the \gls{SN} controller, the client resumes \gls{BP} to update all weights in the head model, thereby completing the training step. This process, illustrated in \cref{algo:Headsplit} (line number 6-9 and 24-29), is iteratively repeated across all epochs until convergence. The device-side computation model represents the \gls{FP} process of the head model. Let $\delta_{d}^{F}(\boldsymbol{\nu})$ denote the computation workload (in \gls{FLOPS}) for one data sample on the $\boldsymbol{\nu}$-th cut layer. This model processes $S_{w}$ samples, resulting in an overall computation workload of $S_{w} \cdot \delta_{d}^{F}(\boldsymbol{\nu})$. The overall computation latency is given by \cite{lin2024efficient}:
\begin{equation}
T_{m_{\tau},b}^{FP} = \dfrac{S_{w}\cdot \kappa \cdot \delta_{d}^{F}(\boldsymbol{\nu})}{\varrho_{m}}, \forall m_{\tau} \in \mathcal{M} ,
\end{equation}
where $\varrho_{m}$ denotes the \gls{CPU} capability of device $m_{\tau}$, and $\kappa$ represents the computing intensity. The communication model represents the transmission of smashed data generated by the head model. Let the size of the smashed data, in bits, corresponding to $S_{w}$ samples on the $\boldsymbol{\nu}$-th cut layer be $S_{w} \Lambda_{s}(\boldsymbol{\nu})$. Considering that $R^{UL}$ is the uplink transmission rate, the smashed data transmission latency is given by \cite{lin2024efficient}: 
\begin{equation}
T_{m_{\tau},b}^{FP} = \dfrac{S_{w} \Lambda_{s}(\boldsymbol{\nu})}{R^{UL}}, \forall m_{\tau} \in \mathcal{M} ,
\end{equation}

\begin{algorithm}
\caption{Split-\gls{iQPTransformer}-in-X Algorithm}
\label{algo:Headsplit}
\begin{algorithmic}[1]
\State \textbf{Input:}  Training data $\mathbf{D}_{\text{train}}$ of $n_q$-th \gls{SN}
\Statex
\State Initialize model $\mathbf{w}_{0}^{m_{i}}, \forall m_{i} \in \mathcal{M}_{n_q}$ with Xavier initialization \cite{datta2020survey}
\For{each epoch ${e\in \{1,2,\dotsc,E\}}$ }
\Statex
\State \textbf{Feedforward Propagation:}
\For{each client $m_{i \in \{1,2,\dotsc,M\}}$ in parallel}
\State \textbf{Head Model}
\State Feedforward propagation $\mathbf{w}_{e}^{m_{i}}$ with input $\mathbf{D}_{\text{train}}^{m_{i}}$
\State Calculate activation function $\mathbf{A}_{e}^{m_{i}}$ at its cut layer
\State send ($\mathbf{A}_{e}^{m_{i}}$) to server
\EndFor
\State \textbf{Body Model:}
\State Concatenate activations:  $\mathbf{A}_{e} = \text{concat}(\mathbf{A}^{m_{1}}_{e} , \dotsc, \mathbf{A}^{m_{M}}_{e})$
\State Feedforward propagation $\mathbf{w}_{e}^{S}$ 
\State Calculate activation function $\mathbf{A}_{e}^{S}$ at its cut layer
\State send ($\mathbf{A}_{e}^{S}$) to the client tail model
\For{each client $m_{i \in \{1,2,\dotsc,M\}}$ in parallel}
\State \textbf{Tail Model:}
\State Feedforward propagation $\mathbf{\theta}_{e}^{m_{i}}$ with input $\mathbf{A}^{S}_{e} $
\State Calculate loss with label $\mathbf{y}_{m_{i}}$ and prediction \textbf{$\hat{\mathbf{y}}_{m_{i}}$}
\State $J \rightarrow \mathcal{L}(\mathbf{y}_{m_{i}},\hat{\mathbf{y}}_{m_{i}})$ \Comment{See \cref{eq:pinball}}
\EndFor
\Statex
\State \textbf{Backward Propagation:}
\For{each client $m_{i \in \{1,2,\dotsc,M\}}$ in parallel}
\State \textbf{Head Model:}
\State Receive $\Big(\dfrac{\partial J}{\partial a{'}^{(L_{s})}}\Big)$ from server
\For{ $i \leftarrow L_{h}$ down to $1$}
\State compute $\Bigg\{\dfrac{\partial J}{\partial \mathbf{w}_{e}^{m_{i}}}, \dfrac{\partial J}{\partial \mathbf{b}_{e}^{m_{i}}} \Bigg\}$
\State update $\mathbf{w}_{e}^{m_{i}},b_{e}^{m_{i}}$
\EndFor
\EndFor
\State \textbf{Body Model:}
\State Receive $\Big(\dfrac{\partial J}{\partial a{'}^{(L_{s})}}\Big)$ from client tail model
\For{ $i \leftarrow L_{s}$ down to $L_{h}+1$}
\State compute $\Bigg\{\dfrac{\partial J}{\partial \mathbf{w}_{e}^{m_{i}}}, \dfrac{\partial J}{\partial \mathbf{b}_{e}^{m_{i}}} \Bigg\}$
\State update $\mathbf{w}_{e}^{m_{i}},\mathbf{b}_{e}^{m_{i}}$
\EndFor
\For{each client $m_{i \in \{1,2,\dotsc,M\}}$ in parallel}
\State \textbf{Tail Model:}
\For{ $i \leftarrow L$ down to $L_{s}+1$}
\State compute $\Bigg\{\dfrac{\partial J}{\partial \hat{\mathbf{y}}_{m_{i}} }, \text{and}~ \dfrac{\partial J}{\partial \mathbf{b}_{e}^{m_{i}}} \Bigg\}$
\State update $\mathbf{w}_{e}^{m_{i}},\mathbf{b}_{e}^{m_{i}}$
\EndFor
\State compute and send to server $(\dfrac{\partial J}{\partial \mathbf{A}(i)})$
\EndFor
\EndFor
\Statex
\end{algorithmic}
\end{algorithm}

\subsubsection{\gls{iQPTransformer} Split Server (Body Model)}
Upon receiving the smashed data from a client head model, the server processes the data using \gls{FP} through its body model. The body model consists of three key components. First, the multivariate attention layer captures dependencies between different \gls{SA} pairs and exploits a cross-correlation map using query, key, and value matrices. This enables the server to incorporate spatial information from other \gls{SA} pairs, improving the model’s understanding of interference dynamics. Second, the \gls{LSTM} layer captures temporal dependencies within each \gls{SA} pair. Finally, layer normalization is applied to stabilize training by normalizing inputs, ensuring consistent activations across layers.  During backward propagation, the server computes gradients up to the cut layer and sends them back to the corresponding client for \gls{BP}. The functionality of the body model is summarized in \cref{algo:Headsplit} (line number 11-15 and 31-36). Since all the smashed data are fed for training the server-side model, the number of the concatenated smashed data samples is $K_m \cdot S_{w}$, and the overall computation workload is $K_m \cdot S_{w} \cdot \delta_{d}^{F}(\boldsymbol{\nu})$. Taking \gls{FP} into account, the overall latency can be given as \cite{lin2024efficient}
\begin{equation}\label{eq:comp_sn}
T_{e} = \dfrac{K_m \cdot S_{w} \cdot \kappa \cdot \delta_{d}^{F}(\boldsymbol{\nu})}{\varrho_{m}}, \forall m_{\tau} \in \mathcal{M}.
\end{equation}
Considering that $R^{UL}$ is the uplink transmission rate, the smashed data transmission latency is given by \cite{lin2024efficient}: 
\begin{equation}
T_{b,m_{\tau}}^{FP} = \dfrac{K_m \cdot S_{w} \Lambda_{s}(\boldsymbol{\nu})}{R^{UL}}, \forall m_{\tau} \in \mathcal{M} ,
\end{equation}

\subsubsection{\gls{iQPTransformer} Split Client (Tail Model)}
Once the client receives the processed data from the server, \gls{FP} continues through the tail model. The tail model is responsible for refining predictions and includes two key components. First, the quantile projection layer predicts dynamic probabilistic thresholds for interference, enabling the model to handle uncertainty quantification in a distributed fashion. Following forward propagation, the client calculates the training loss and performs \gls{BP} through the tail model, sending gradients back to the server. This process, detailed in \cref{algo:Headsplit}~(line numbers 17-20 and 38-43), is repeated across all epochs. Following a similar formulation as the head model, the computation latency for \gls{FP} can be expressed as:
\begin{equation}
T_{b,m_{\tau}}^{FP} = \dfrac{S_{w} \cdot \kappa \cdot \delta_{d}^{F}(\boldsymbol{\nu})}{\varrho_{m}}, \forall m_{\tau} \in \mathcal{M} ,
\end{equation}
where $\delta_{d}^{B}(\boldsymbol{\nu})$ represents the computation workload (in \gls{FLOPS}) for one data sample during the \gls{BP} process at the $\boldsymbol{\nu}$-th cut layer. Overall, the latency for execution of split learning for training for forward propagation~(considering the auto-regressive nature of transformer) can be given as
\begin{equation}
    T_{\text{split}} = \underbrace{T_{m_{\tau},b}^{FP} +T_{e}+T_{b,m_{\tau}}^{FP}}_{\text{Computation Time}}+ \underbrace{T_{b,m_{\tau}}^{FP}+T_{m_{\tau},b}^{FP}}_{\text{Communication Time}}.
\end{equation}
Based on our assumption regarding the \gls{CPU} capability $f_{k}$, computational intensity $\kappa$, and computation overload $\delta_{d}^{B}$ of \gls{SN} controller and \gls{SA} pair, the model is distributed as shown in \cref{fig:lstm_split}, considering these parameters are fixed and known to the designer \footnote{Meanwhile, utilizing dynamic constraint as an optimization problem to probabilistically predict interference \cref{eq:prob_eq_1} remains an open topic for future research.}. For the Split-\gls{iQPTransformer}, the computational complexity remains \( \mathcal{O}(S_w \cdot M^2 \cdot d_e + S_w \cdot M \cdot H^2) \) but with an additional \( \mathcal{O}(M \cdot d_e) \) term, which accounts for the communication overhead from transmitting intermediate activations between the distributed components. Despite this added communication cost, the overall computational complexity remains identical to the centralized model, ensuring scalability while distributing the computational burden between the \gls{SA} pair and the \gls{SN} controller. During training, the loss is monitored, and the reliability of the predictions is evaluated by measuring the coverage width and probability and its impact on achieved \gls{BLER} with resource allocation.

\section{Numerical Results}
\label{sec:simulations}
\subsection{Baseline Methods}
The following baseline methods are considered for comparison in this work:

\subsubsection{Genie RA:} This baseline selects the channel usages based on the target \gls{BLER} ${\varepsilon}_{\text{target}}$ under the assumption that a \gls{SA} pair's \gls{IPV} is predicted perfectly at the \gls{SN} controller. Note that Genie \gls{RA} is impractical due to the assumption of perfect \gls{IPV} prediction \cite{mahmood2020predictive}. 

\subsubsection{Moving-average (MA) Predictor:} In this baseline, the \gls{IPV} at every $k$ is obtained as a weighted-sum of \glspl{IPV} estimated at \glspl{TTI} $k-1$ and $k-2$ \cite{mahmood2020predictive}. This method gives the efficacy and impacts of utilizing average prediction in \gls{HRLLC} use-case. The computational complexity of this predictor is \(\mathcal{O}(1)\).
\subsubsection{Wiener Predictor:} Interference power values can be considered as a short-term stationary process like \gls{SINR} as mentioned in \cite{xu2013improving}. This predictor uses sample interference correlation instead of the autocorrelation function. We have predicted interference using order $S_{w}$ Wiener filter \cite{pramesh2024int}. The performance of predicted interference depends on the accuracy of the sample interference correlation. For one-step prediction, the resulting equation is identical to the Yule-Walker equations which are used to solve the autoregressive~($S_{w}$) process \cite{kay1993fundamentals}. This method implicitly utilizes second-order statistics of interference. The overall computational complexity of this predictor with direct matrix inversion is \(\mathcal{O}(S_w^3)\).
\subsection{Performance Metrics}
\subsubsection{Coverage Probability}
The coverage probability measures the ability of the proposed predictor to ensure that the measured interference power lies within the predicted interval. In order to provide reliable performance, the predicted interference should be equal to or greater than the Genie/true interference power value. The equation relating to coverage probability for each \gls{SA} pair is defined below \cite{jensen2022ensemble}:
\begin{equation}
    \dfrac{\sum_{i = 1}^{L_{\text{test}}} M_{i}}{L_{\text{test}}} , M_{i} = \begin{cases}
      1 & \Tilde{I}_{m_\tau,n_q}(t) \leq \hat{I}_{m_\tau,n_q}(t)\\
      0 & \text{otherwise}
    \end{cases}  
\end{equation}
where $\hat{I}_{m_\tau,n_q}(t)$ is the predicted interference power value. This metric is often referred to as the reliability of the predictor and has been used to define the overall reliability for link adaptation in \cite{brighente2022interference}.

\subsubsection{Coverage Width}
The coverage  width or sharpness of the interference predictor can be defined as follows \cite{jensen2022ensemble}:
\begin{equation}
    \dfrac{1}{L_{\text{test}}}\sum_{i = 1}^{L_{\text{test}}}  |\hat{I}_{m_\tau,n_q}(t)-\Tilde{I}_{m_\tau,n_q}(t)|
\end{equation}
Our proposed learning method should aim to achieve a higher coverage probability and a lower coverage width, ensuring the target \gls{BLER} is achieved.

\subsubsection{Target vs Achieved \gls{BLER}}
This performance metric highlights the discrepancy between the achieved \gls{BLER} and the selected channel usage according to the predicted interference and the target \glspl{BLER}. Firstly, the channel usage is selected using predicted interference and the target \gls{BLER}. Later, the result obtained \gls{BLER} is computed using the same channel usages but with actual interference for accuracy evaluation.

\subsection{Simulation Setup}
We consider two mobility models to evaluate the performance of the proposed technique. The first model captures the movement of densely deployed robots within a fixed area corresponding to a single production module, while the second represents a more complex mobility model in which each \gls{SN} moves along designated alleys within an indoor factory layout, as described in \cite{5G-ACIA}. 

First, we evaluate the performance of the \glspl{SN}, modeled as a densely deployed network operating concurrently in a factory to perform collaborative tasks. Specifically, we consider the deployment of 16 \glspl{SN} in a $25 \times 25~\mathrm{m}^2$ region, where each \gls{SN} moves at a speed of $2~\mathrm{m/s}$ in a random direction, with ${\vartheta \sim U(0, 2\pi)}$. Each \gls{SN} consists of $M$ \gls{SA} pairs, which are distributed within a radius $r = 2$ from the center of the \gls{SN}, following a binomial point process. To enhance the realism of this mobility model, we integrate additional features such as collision avoidance—a common capability of autonomous robots—and restrict their movement to a confined area. The deployment parameters are summarized in \cref{table:1}, and the channel parameters, based on the \gls{3GPP} \cite{3gpp2020study}, are also detailed in \cref{table:1}. The estimated interference is restructured as described in \cref{sec:reliableprediction}, in accordance with the exchangeability property. Subsequently, the proposed model—incorporating the \gls{iQPTransformer} and split-\gls{iQPTransformer} parameters—is trained for 300 epochs, as outlined in \cref{table:1}.

\begin{table}
\centering
\caption{ Simulation Parameters}
\label{table:1}
\resizebox{\linewidth}{!}{
\begin{tabular}{ll} \hline
\textbf{Parameter}                                          & \textbf{Value}~  \\ \hline
Deployment Parameter                                       \\ \hline
Number of \glspl{SN} , $|\mathcal{N}|$                                   & 16               \\ 
Number of sensor-actuator pair, $|\mathcal{M}|$                           & \{4,8,12,16\}                \\
Interfering \glspl{SN}, $|\mathcal{K}|$                           & 5                \\
Deployment density~(\glspl{SN}/$km^2$) & 25600            \\ 
Mobility Model                                              & \begin{tabular}[c]{@{}l@{}}Robots working on the production line \\ with RDMM moving along predefined paths.\end{tabular}             \\
Cell Radius, r                                    & 2 [m]            \\
Velocity, $v$                                             & 2 [m/s]          \\
Minimum distance, $d$                                     & 3 [m]            \\ \hline
Channel Parameter         \\ \hline
Carrier frequency                             & $6$ GHz         \\
Number of sub-bands, $\Omega$                           & $4$                \\
Frequency reuse                            & $1/4$                \\
Pathloss               & \gls{3GPP} InF-DL \cite{3gpp2018study}            \\
Shadow fading std. deviation                            & $4$ dB(LOS)$~|~$$7.2$ dB(NLOS)        \\
Decorrelation distance,           & 10 m            \\
Doppler frequency, $f_{d_n}$                   &  $80~\text{Hz}$ \\
Transmit Power          &  $0$ dBW          \\
\gls{TX} cycle duration &  $1$ ms \\
Packet size & $200$ bits \\
\hline
\gls{iQPTransformer} Parameters                                          \\ \hline
Activation Function          & tanh            \\ 
Loss Function          & Pinball Loss function            \\
Learning Rate and Training epoch          & 0.001  and 300          \\
Encoder Layer          & 2 \\
Embedding hidden units and dropout         & 256 and 0.1\\
number of heads for attention, hidden units, and dropout & 8, 256 and 0.1 \\
Number of LSTM layers and hidden units per layer          & 1 and 256           \\
Optimizer          & Adam            \\
Weight Initialization          & Xavier            \\
Batch Size          & 128            \\
Quantile assocated with \gls{QR}, $1-\alpha$         & 0.95            \\
\gls{EVT}-quantile, $1-\varsigma$         & 0.5            \\
Confidence level associated with \gls{QR}, $1-\beta$         & 0.95            \\
Train, calibration, and test data size        & 7000, 1000, 2000            \\
\hline
\end{tabular}}
\end{table}

\subsection{Performance Analysis}

\begin{table*}[]
\centering
\caption{Normalized Coverage width and probability of baseline and proposed predictor}
\label{table:cov_wid_probab}
\begin{tabular}{|l|l|lllll|lllll|}
\hline
\rowcolor[HTML]{C0C0C0} 
\cellcolor[HTML]{C0C0C0}                   & \multicolumn{1}{c|}{\cellcolor[HTML]{C0C0C0}}                            & \multicolumn{5}{l|}{\cellcolor[HTML]{C0C0C0}Coverage Probability}                                                                                                                                                                                 & \multicolumn{5}{l|}{\cellcolor[HTML]{C0C0C0}Normalized Coverage Width}                                                                                                                                                                            \\ \cline{3-12} 
\rowcolor[HTML]{C0C0C0} 
\multirow{-2}{*}{\cellcolor[HTML]{C0C0C0}} & \multicolumn{1}{c|}{\multirow{-2}{*}{\cellcolor[HTML]{C0C0C0}Predictor}} & \multicolumn{1}{l|}{\cellcolor[HTML]{C0C0C0}SA1}   & \multicolumn{1}{l|}{\cellcolor[HTML]{C0C0C0}SA2}   & \multicolumn{1}{l|}{\cellcolor[HTML]{C0C0C0}SA3}   & \multicolumn{1}{l|}{\cellcolor[HTML]{C0C0C0}SA4}   & Average                           & \multicolumn{1}{l|}{\cellcolor[HTML]{C0C0C0}SA1}   & \multicolumn{1}{l|}{\cellcolor[HTML]{C0C0C0}SA2}   & \multicolumn{1}{l|}{\cellcolor[HTML]{C0C0C0}SA3}   & \multicolumn{1}{l|}{\cellcolor[HTML]{C0C0C0}SA4}   & Average                           \\ \hline
1                                          & Moving Average                                                           & \multicolumn{1}{l|}{0.405}                         & \multicolumn{1}{l|}{0.277}                         & \multicolumn{1}{l|}{0.152}                         & \multicolumn{1}{l|}{0.564}                         & 0.350                         & \multicolumn{1}{l|}{0.131}                         & \multicolumn{1}{l|}{0.163}                         & \multicolumn{1}{l|}{0.208}                         & \multicolumn{1}{l|}{0.139}                         & 0.160                         \\ \hline
2                                          & Wiener                                                                   & \multicolumn{1}{l|}{0.409}                         & \multicolumn{1}{l|}{0.366}                         & \multicolumn{1}{l|}{0.188}                         & \multicolumn{1}{l|}{0.549}                         & 0.378                         & \multicolumn{1}{l|}{0.153}                         & \multicolumn{1}{l|}{0.202}                         & \multicolumn{1}{l|}{0.210}                         & \multicolumn{1}{l|}{0.202}                         & 0.192                         \\ \hline
3                                          & \gls{iQPTransformer}                                                            & \multicolumn{1}{l|}{0.948}                         & \multicolumn{1}{l|}{0.958}                         & \multicolumn{1}{l|}{0.938}                         & \multicolumn{1}{l|}{0.935}                         & 0.945                         & \multicolumn{1}{l|}{\textbf{0.052}} & \multicolumn{1}{l|}{\textbf{0.058}} & \multicolumn{1}{l|}{\textbf{0.057}} & \multicolumn{1}{l|}{\textbf{0.067}} & \textbf{0.058} \\ \hline
4                                          & \gls{iQPTransformer}\_split                                                     & \multicolumn{1}{l|}{0.951}                         & \multicolumn{1}{l|}{0.922}                         & \multicolumn{1}{l|}{0.945}                         & \multicolumn{1}{l|}{0.936}                         & 0.938                         & \multicolumn{1}{l|}{0.058}                         & \multicolumn{1}{l|}{0.064}                         & \multicolumn{1}{l|}{0.068}                         & \multicolumn{1}{l|}{0.072}                         & 0.065                         \\ \hline
5                                          & \gls{EVT} \_\gls{iQPTransformer}                                                      & \multicolumn{1}{l|}{0.974}                         & \multicolumn{1}{l|}{0.971}                         & \multicolumn{1}{l|}{0.960}                         & \multicolumn{1}{l|}{0.958}                         & 0.966                         & \multicolumn{1}{l|}{0.057}                         & \multicolumn{1}{l|}{0.062}                         & \multicolumn{1}{l|}{0.060}                         & \multicolumn{1}{l|}{0.071}                         & 0.062                         \\ \hline
6                                          & \gls{EVT} \_\gls{iQPTransformer}\_split                                                & \multicolumn{1}{l|}{0.971}                         & \multicolumn{1}{l|}{0.948}                         & \multicolumn{1}{l|}{0.964}                         & \multicolumn{1}{l|}{0.952}                         & 0.954                         & \multicolumn{1}{l|}{0.063}                         & \multicolumn{1}{l|}{0.069}                         & \multicolumn{1}{l|}{0.072}                         & \multicolumn{1}{l|}{0.076}                         & 0.070                         \\ \hline
7                                          & CEVT\_\gls{iQPTransformer}                                                      & \multicolumn{1}{l|}{\textbf{0.998}} & \multicolumn{1}{l|}{\textbf{0.998}} & \multicolumn{1}{l|}{\textbf{0.998}} & \multicolumn{1}{l|}{\textbf{0.997}} & \textbf{0.998} & \multicolumn{1}{l|}{0.095}                         & \multicolumn{1}{l|}{0.108}                         & \multicolumn{1}{l|}{0.100}                         & \multicolumn{1}{l|}{0.115}                         & 0.107                         \\ \hline
8                                          & CEVT\_\gls{iQPTransformer}\_split                                               & \multicolumn{1}{l|}{\textbf{0.998}}                         & \multicolumn{1}{l|}{0.993}                         & \multicolumn{1}{l|}{0.997}                         & \multicolumn{1}{l|}{\textbf{0.997}}                         & \textbf{0.998}                         & \multicolumn{1}{l|}{0.122}                         & \multicolumn{1}{l|}{0.105}                         & \multicolumn{1}{l|}{0.132}                         & \multicolumn{1}{l|}{0.130}                         & 0.122                         \\ \hline
\end{tabular}
\end{table*}
We have evaluated the performance of the proposed interference prediction technique against three baseline schemes. First, we evaluate its performance with Bernoulli isochronous traffic, followed by an analysis with push- and pull-based traffic.
\begin{figure*}[!hbt]
    \centering
    \includegraphics[width=0.95\linewidth]{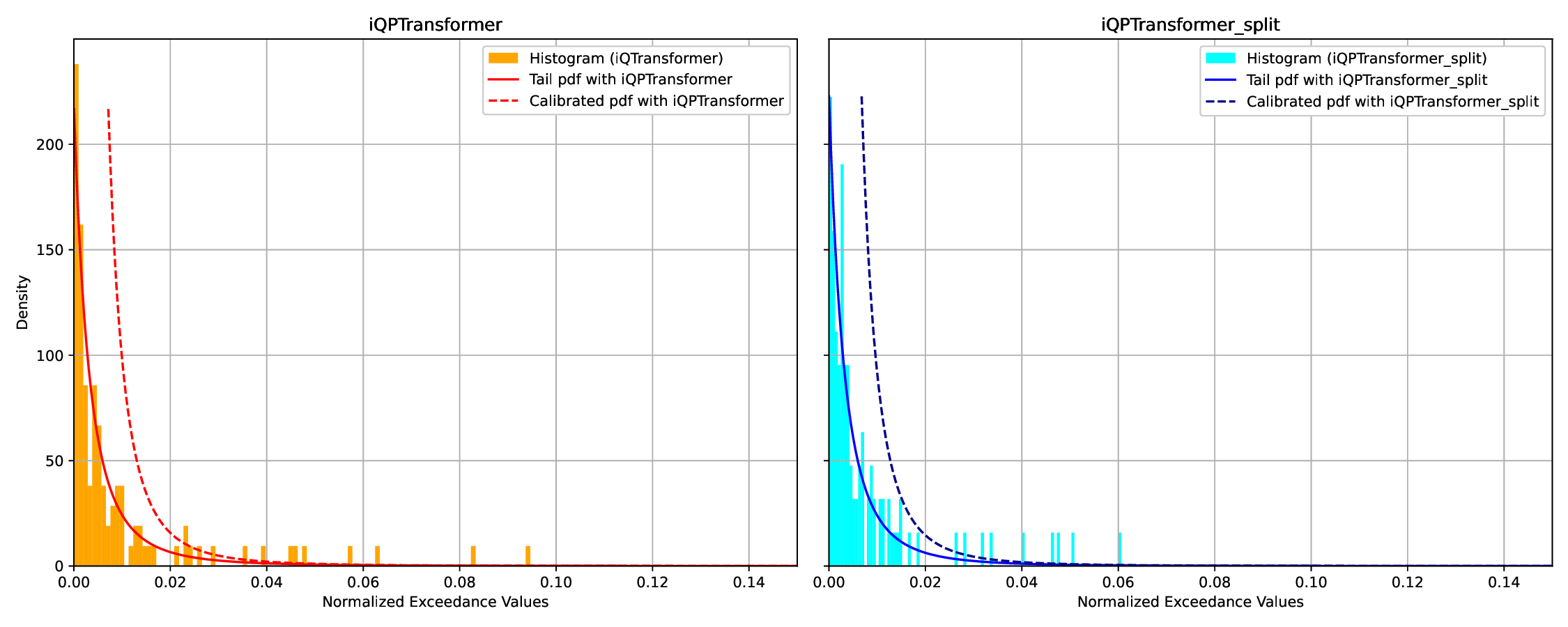}
    \caption{Calibrated tail \gls{pdf}  of interference}
    \label{fig:tail_pdf_fig}
\end{figure*}
As shown in \cref{table:cov_wid_probab}, the moving average predictor demonstrates inadequate performance, achieving an average coverage probability of only 0.35 due to two inherent limitations: its inability to capture local interference dynamics and its lack of support for tail statistics. These known drawbacks highlight the need for predictors that move beyond mean-based prediction to ensure reliability in dynamic wireless environments with sporadic traffic. The Wiener predictor partially addresses this by leveraging second-order statistics to track interference dynamics under varying traffic and channel statistics. In contrast, the proposed \gls{iQPTransformer} captures highly non-linear interference dynamics while quantifying uncertainty using tail statistics at the $1 - \alpha = 0.95$ level (heuristically selected). The distributed \gls{iQPTransformer}-split exhibits only a maximum degradation of 0.036 and an average degradation of 0.07 in coverage probability across all \gls{SA} pairs, indicating minimal performance disparity compared to its centralized counterpart. The calibrated tail \gls{pdf} is illustrated in \cref{fig:tail_pdf_fig}, which clearly shows that the tail of the exceedance values is shifted by 0.0071 and 0.00676 for the centralized and split \gls{iQPTransformer}, respectively, by accounting for additional uncertainties in the design of a risk-aware prediction scheme. By utilizing calibrated tail \glspl{pdf} at the $1 - \varsigma=  0.5$ quantile, we achieve an average coverage probability of 0.998, with minimum values of 0.997 and 0.993 for the centralized and distributed approaches, respectively. These coverage probabilities are achieved with normalized average coverage widths of 2.65 dB and 2.73 dB, both lower than those of the baseline methods. The calibrated tail \gls{pdf} also enables flexible selection of $\varsigma,~\beta$ for resource allocation while ensuring statistical reliability. When $\varsigma = 1$, the value converges to the calibrated interference threshold, and when $\varsigma = 0$, it approaches the tail endpoint. This flexibility allows both \gls{iQPTransformer} variants to produce slightly conservative predictions that track interference dynamics while accounting for extreme events, introducing robust and risk-aware predictive interference management in \gls{HRLLC} demanding \gls{SN}.

\begin{figure*}
    \centering
    \includegraphics[width=0.95\linewidth]{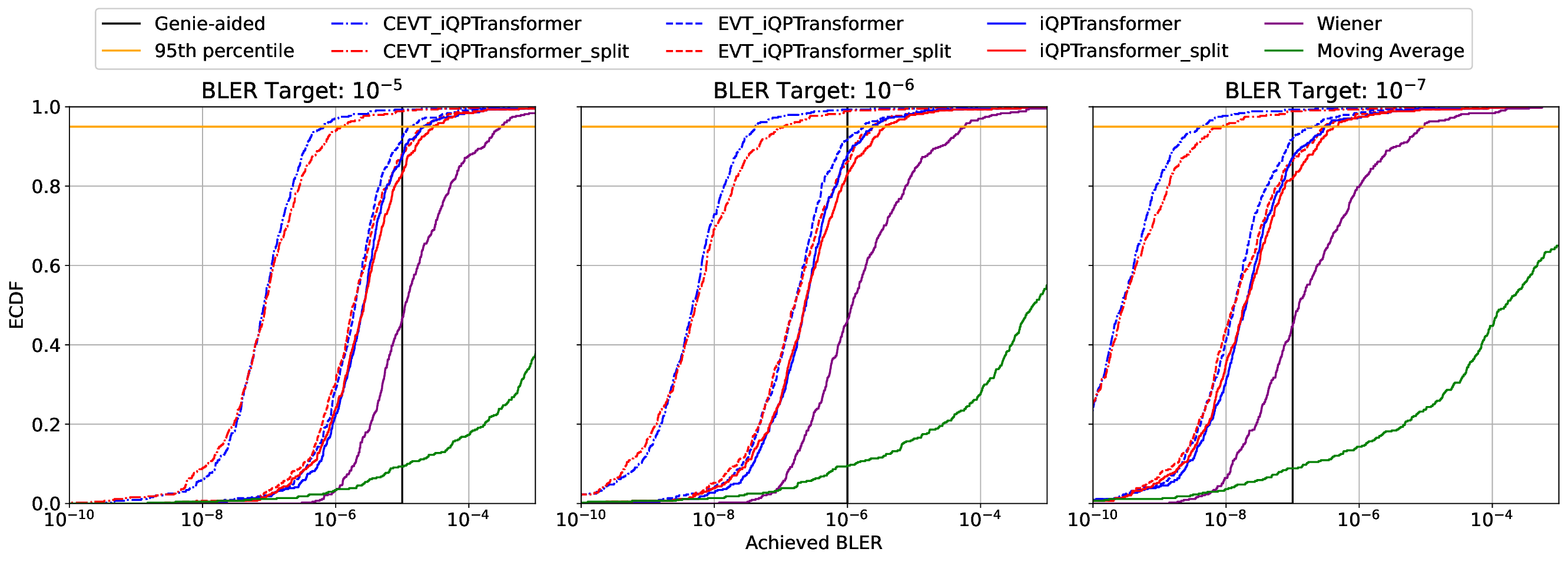}
    \caption{Achieved \gls{BLER} vs target \gls{BLER} in the range of \gls{HRLLC} requirements}
    \label{fig:achieved_bler}
\end{figure*}

\subsubsection{Impact of scalability}

We evaluate the impact of scaling the number of \gls{SA} pairs on prediction performance to assess the efficacy of the proposed techniques. The number of \gls{SA} pairs is increased from 4 to 16 within the $n_q$th \gls{SN}. As shown in \cref{tab:scaling_ncw} and \cref{tab:scaling_cp}, traditional methods such as the moving average and Wiener predictors exhibit higher normalized coverage widths and lower coverage probabilities—often referred to as predictor reliability. In contrast, the proposed \gls{iQPTransformer} and its variants (\gls{iQPTransformer}-split and CPEVT\_\gls{iQPTransformer}) achieve significantly better results in terms of both coverage width and probability. Specifically, the \gls{iQPTransformer} maintains lower normalized coverage widths (ranging from 0.035 to 0.039 on average) while achieving average coverage probabilities exceeding 0.946. The inclusion of calibrated tail statistics in the CPEVT\_\gls{iQPTransformer} further improves performance, reaching near-optimal coverage probabilities (above 0.997) with only marginal increases in coverage width. These results highlight the scalability and robustness of both centralized and \gls{iQPTransformer} and split-\gls{iQPTransformer} methods, making them suitable for multi-\gls{SA} pair deployments. While coverage probability and width provide a high-level overview of interference prediction, we further discuss how they impact system-level performance—particularly the achieved \gls{BLER}.

\begin{table*}[]
\centering
\caption{Normalized coverage width with scaling number of \gls{SA} pairs}
\label{tab:scaling_ncw}
\begin{tabular}{|l|llllllll|}
\hline
\multicolumn{1}{|c|}{\multirow{3}{*}{Methods}} & \multicolumn{8}{c|}{Normalized Coverage Width}                                                                                                                                                                                   \\ \cline{2-9} 
\multicolumn{1}{|c|}{}                         & \multicolumn{2}{l|}{4}                                      & \multicolumn{2}{l|}{8}                                      & \multicolumn{2}{l|}{12}                                     & \multicolumn{2}{l|}{16}                \\ \cline{2-9} 
\multicolumn{1}{|c|}{}                         & \multicolumn{1}{l|}{Maximum} & \multicolumn{1}{l|}{Average} & \multicolumn{1}{l|}{Maximum} & \multicolumn{1}{l|}{Average} & \multicolumn{1}{l|}{Maximum} & \multicolumn{1}{l|}{Average} & \multicolumn{1}{l|}{Maximum} & Average \\ \hline
Moving Average                                 & \multicolumn{1}{l|}{0.149}   & \multicolumn{1}{l|}{0.123}   & \multicolumn{1}{l|}{0.149}   & \multicolumn{1}{l|}{0.118}   & \multicolumn{1}{l|}{0.163}   & \multicolumn{1}{l|}{0.118}   & \multicolumn{1}{l|}{0.163}   & 0.117   \\ \hline
Wiener                                         & \multicolumn{1}{l|}{0.189}   & \multicolumn{1}{l|}{0.167}   & \multicolumn{1}{l|}{0.203}   & \multicolumn{1}{l|}{0.165}   & \multicolumn{1}{l|}{0.203}   & \multicolumn{1}{l|}{0.16}    & \multicolumn{1}{l|}{0.203}   & 0.16    \\ \hline
iQPTransformer                                  & \multicolumn{1}{l|}{\textbf{0.043}}   & \multicolumn{1}{l|}{\textbf{0.039}}   & \multicolumn{1}{l|}{\textbf{0.041}}   & \multicolumn{1}{l|}{\textbf{0.036}}   & \multicolumn{1}{l|}{\textbf{0.049}}   & \multicolumn{1}{l|}{\textbf{0.038}}   & \multicolumn{1}{l|}{\textbf{0.046}}   & \textbf{0.035}   \\ \hline
CEVT\_iQPTransformer                            & \multicolumn{1}{l|}{0.063}   & \multicolumn{1}{l|}{0.058}   & \multicolumn{1}{l|}{0.066}   & \multicolumn{1}{l|}{0.058}   & \multicolumn{1}{l|}{0.097}   & \multicolumn{1}{l|}{0.064}   & \multicolumn{1}{l|}{0.085}   & 0.059   \\ \hline
iQPTransformer\_split                           & \multicolumn{1}{l|}{0.049}   & \multicolumn{1}{l|}{0.044}   & \multicolumn{1}{l|}{0.052}   & \multicolumn{1}{l|}{0.044}   & \multicolumn{1}{l|}{0.06}    & \multicolumn{1}{l|}{0.046}   & \multicolumn{1}{l|}{0.062}   & 0.045   \\ \hline
CEVT\_iQPTransformer\_split                     & \multicolumn{1}{l|}{0.129}   & \multicolumn{1}{l|}{0.09}    & \multicolumn{1}{l|}{0.102}   & \multicolumn{1}{l|}{0.081}   & \multicolumn{1}{l|}{0.133}   & \multicolumn{1}{l|}{0.084}   & \multicolumn{1}{l|}{0.136}   & 0.08    \\ \hline
\end{tabular}
\end{table*}

% Please add the following required packages to your document preamble:
% \usepackage{multirow}
\begin{table*}[]
\caption{Coverage probability with scaling number of \gls{SA} pairs}
\centering
\label{tab:scaling_cp}
\begin{tabular}{|l|llllllll|}
\hline
\multicolumn{1}{|c|}{\multirow{3}{*}{Methods}} & \multicolumn{8}{c|}{Coverage Probability}                                                                                                                                                                                \\ \cline{2-9} 
\multicolumn{1}{|c|}{}                         & \multicolumn{2}{l|}{4}                                    & \multicolumn{2}{l|}{8}                                    & \multicolumn{2}{l|}{12}                                   & \multicolumn{2}{l|}{16}              \\ \cline{2-9} 
\multicolumn{1}{|c|}{}                         & \multicolumn{1}{l|}{Worst} & \multicolumn{1}{l|}{Average} & \multicolumn{1}{l|}{Worst} & \multicolumn{1}{l|}{Average} & \multicolumn{1}{l|}{Worst} & \multicolumn{1}{l|}{Average} & \multicolumn{1}{l|}{Worst} & Average \\ \hline
Moving Average                                 & \multicolumn{1}{l|}{0.537} & \multicolumn{1}{l|}{0.573}   & \multicolumn{1}{l|}{0.537} & \multicolumn{1}{l|}{0.581}   & \multicolumn{1}{l|}{0.537} & \multicolumn{1}{l|}{0.583}   & \multicolumn{1}{l|}{0.537} & 0.586   \\ \hline
Wiener                                         & \multicolumn{1}{l|}{0.491} & \multicolumn{1}{l|}{0.506}   & \multicolumn{1}{l|}{0.491} & \multicolumn{1}{l|}{0.516}   & \multicolumn{1}{l|}{0.491} & \multicolumn{1}{l|}{0.517}   & \multicolumn{1}{l|}{0.491} & 0.517   \\ \hline
iQPTransformer                                  & \multicolumn{1}{l|}{0.946} & \multicolumn{1}{l|}{0.947}   & \multicolumn{1}{l|}{0.928} & \multicolumn{1}{l|}{0.95}    & \multicolumn{1}{l|}{0.959} & \multicolumn{1}{l|}{0.968}   & \multicolumn{1}{l|}{0.952} & 0.962   \\ \hline
CEVT\_iQPTransformer                            & \multicolumn{1}{l|}{0.995} & \multicolumn{1}{l|}{0.997}   & \multicolumn{1}{l|}{\textbf{0.995}} & \multicolumn{1}{l|}{\textbf{0.998}}   & \multicolumn{1}{l|}{\textbf{0.996}} & \multicolumn{1}{l|}{\textbf{0.998}}   & \multicolumn{1}{l|}{\textbf{0.995}} & \textbf{0.998}   \\ \hline
iQPTransformer\_split                           & \multicolumn{1}{l|}{0.945} & \multicolumn{1}{l|}{0.932}   & \multicolumn{1}{l|}{0.932} & \multicolumn{1}{l|}{0.946}   & \multicolumn{1}{l|}{0.926} & \multicolumn{1}{l|}{0.947}   & \multicolumn{1}{l|}{0.932} & 0.946   \\ \hline
CEVT\_iQPTransformer\_split                     & \multicolumn{1}{l|}{\textbf{0.995}} & \multicolumn{1}{l|}{\textbf{0.998}}   & \multicolumn{1}{l|}{0.992} & \multicolumn{1}{l|}{0.997}   & \multicolumn{1}{l|}{0.994} & \multicolumn{1}{l|}{0.997}   & \multicolumn{1}{l|}{0.988} & 0.996   \\ \hline
\end{tabular}
\end{table*}

\subsubsection{Performance within \gls{RA} framework}

\cref{fig:achieved_bler} illustrates the performance comparison of interference prediction techniques in terms of the achieved \gls{BLER} with target \gls{BLER} within the \gls{HRLLC} range (\(10^{-5}\), \(10^{-6}\), and \(10^{-7}\)). The results show that the moving average predictor achieves less than the 20th percentile of target \glspl{BLER}, demonstrating its inability to meet the stringent requirements in the presence of sporadic traffic with non-linear interference dynamics. Although the Wiener predictor can track interference variability by utilizing second-order sample statistics, its performance remains suboptimal, achieving only around the 50th percentile for high target \glspl{BLER}—inadequate for \gls{HRLLC} applications. These limitations stem from their inability to capture interference dynamics and to quantify the associated uncertainties, especially under sporadic traffic and additional sources of randomness. In contrast, the proposed \gls{iQPTransformer} and its variants, such as CEVT\_iQPTransformer and CPEVT\_\gls{iQPTransformer}\_split, demonstrate superior performance. Both proposed methods consistently achieve \gls{BLER} values exceeding the 95th percentile for \gls{BLER} targets utilizing calibrated tail statistics. This directly reflects our consideration that the coverage threshold $1-\beta$ is $95$-th quantile assisting to achieve the 95th percentile of the target \gls{BLER}. By incorporating a calibrated tail \gls{pdf}, the model accurately captures interference extremes, enabling risk-aware and reliable performance in the hyper-reliable target regime. This enhanced reliability comes with an additional channel usage of approximately 10–20\%, representing a deliberate trade-off that remains more efficient than baseline methods.

\begin{figure}
    \centering
    \includegraphics[width=0.95\linewidth]{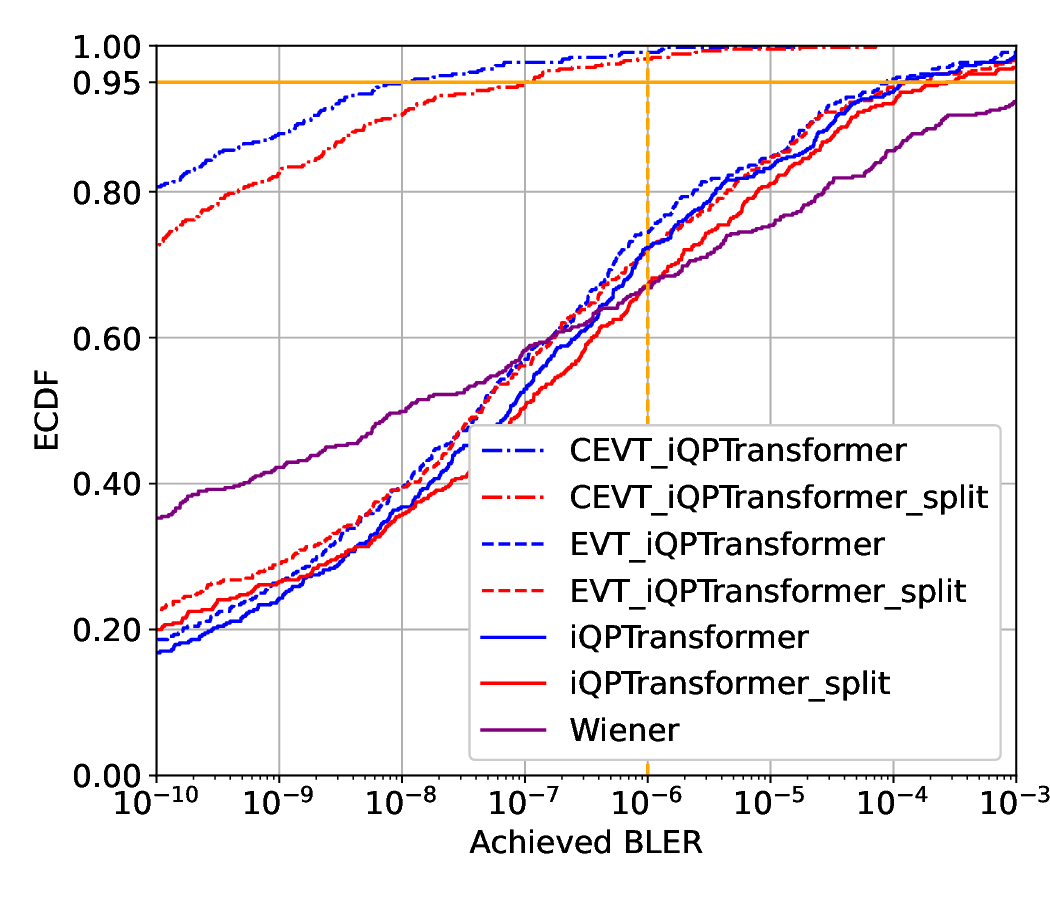}
    \caption{Achieved \gls{BLER} for push-pull-based traffic}
    \label{fig:achieved_bler_push}
\end{figure}

In the case of push- and pull-based traffic, a pool of $ M=6$ \gls{SA} pairs is considered. Two \gls{SA} pairs are scheduled with pull-based deterministic traffic in the first two slots. Among the remaining four \gls{SA} pairs, any two of them transmit packets randomly based on context, illustrating highly unpredictable push-based traffic with a traffic intensity of \(\lambda = 5\).  As a result, the impact of cross-traffic inference between interference power values of \gls{SA} pairs is introduced, stemming from the push- and pull-based traffic as perceived by the \gls{SN} controller.
The \gls{iQPTransformer} achieves high accuracies of 0.957 and 0.947 under pull-based traffic. However, these accuracies decrease to 0.894 and 0.895 under push-based traffic, reflecting the impact of increased traffic unpredictability. Similarly, the \gls{iQPTransformer}-split exhibits a comparable trend, with accuracies of 0.954 and 0.936 under pull-based traffic, which drop to 0.866 and 0.880 under push-based traffic. This consistent decline underscores the challenges introduced by higher traffic randomness. This indicates that, despite slot-level cross-traffic interference dynamics, our proposed inverted token design enables effective inference without negatively impacting performance for other \gls{SA} pairs scheduled in other slots. The utilization of calibrated tail \gls{pdf}, which accounts for those uncertainties, directly helps achieve the 95th-percentile target \gls{BLER} of $10^{-6}$, as shown in \cref{fig:achieved_bler_push}. However, this improvement comes at the cost of increased resource usage—approximately 30–35\% more for pull-based traffic and up to 133\% more for push-based traffic. These results highlight the trade-off between reliability and resource usage in complex traffic scenarios. While baseline methods require 95–125\% more resources with poor \gls{BLER} performance, our risk-aware approach—with the calibrated tail \gls{pdf}—enables reliable predictive resource allocation.  Hence, it can be a potential alternative to naive retransmissions by aligning \gls{RA} decisions with predefined reliability targets, even under sporadic and cross-traffic conditions. Additionally, flexible tuning of the quantile threshold $1 - \alpha$, calibration level $1 - \beta$, and the $1 - \varsigma$ quantile in the \gls{EVT} framework supports efficient resource use across heterogeneous \glspl{SN} based on target \gls{BLER} for diverse use cases.

\subsubsection{Mobility model designed on 5G-ACIA indoor factory layout}
\begin{figure}
    \centering
    \includegraphics[width=0.95\linewidth]{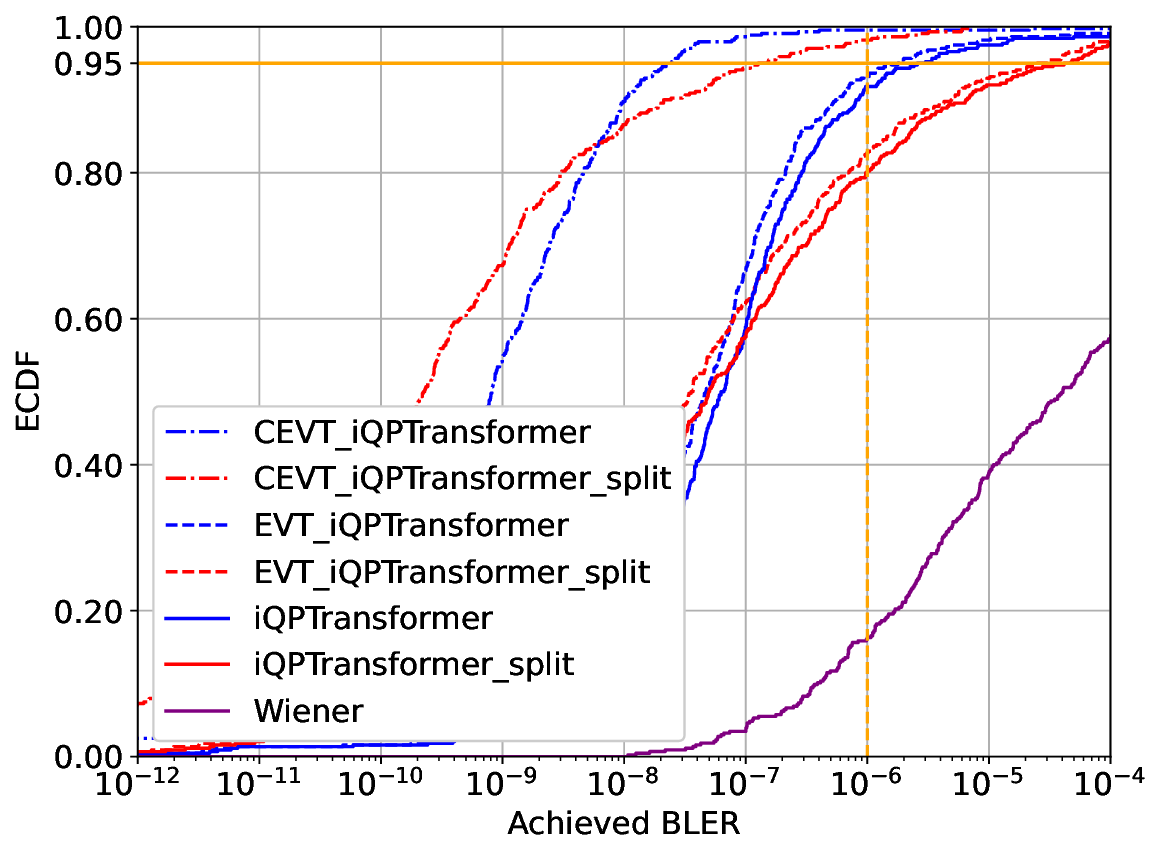}
    \caption{Achieved \gls{BLER} for 5G-ACIA referred indoor factory layout.}
    \label{fig:achieved_bler_new_mm}
\end{figure}

\begin{figure}
    \centering
    \includegraphics[width=0.95\linewidth]{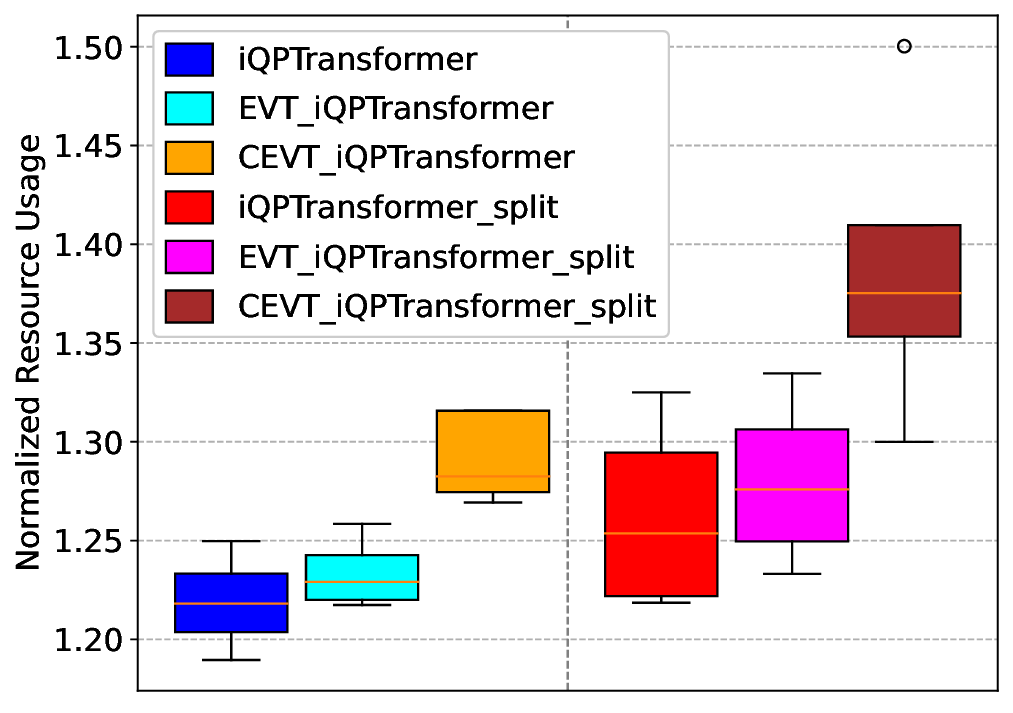}
    \caption{Normalized resource usage}
    \label{fig:normalized_resource_usage}
\end{figure}

Until this point, our evaluation has been based on the \gls{RDMM} reflecting multiple \glspl{SN} operating in the factory hall. However, to further assess the applicability of our proposed algorithm in a more realistic setting, we also evaluate its performance using the indoor factory layout of the 5G Alliance for Connected Industries and Automation (5G-ACIA). In this scenario, we consider 120 \gls{SN} pairs deployed within a $180 \times 90m²$ indoor factory floor, where the nodes move along designated alleys, reflecting complex industrial \gls{SN} mobility. Each \gls{SN} communicates over shared subbands, and for interference evaluation, we consider the 10 nearest \glspl{SN} using the same subband. As shown in \cref{fig:achieved_bler_new_mm}, our proposed hybrid interference prediction technique, enhanced with calibrated tail statistics, enables the system to achieve the target \gls{BLER} even beyond the 95th percentile of interference conditions. \Cref{fig:normalized_resource_usage} illustrates the normalized resource usage, which ranges from $1.2 \times$ to $1.5 \times$. Our proposed predictive method remains more resource-efficient compared to naive blind repetition and packet retransmission, while maintaining the desired \gls{BLER} performance.

These findings emphasize the limitations of traditional methods, not only Moving Average and Wiener predictors, but also \gls{iQPTransformer}-based techniques. The inclusion of calibrated tail statistics within such hybrid statistical and \gls{ML}-based approaches helps to enhance prediction accuracy and adaptability to dynamic interference scenarios, either in a centralized or distributed manner. The proposed methods capture interference variability, enabling reliable communication and efficient resource utilization in dense, interference-prone \gls{SN} deployments to design hyper-reliable and risk-aware \gls{RA}.
\section{Conclusion}
\label{sec:conclusion}
We propose a hybrid framework that leverages interference thresholds predicted by the \gls{iQPTransformer}, combining \gls{EVT} with conformalized quantile regression to model calibrated tail \gls{pdf} for risk-aware and reliable \gls{RA} in hyper-densely deployed 6G industrial \glspl{SN}. We demonstrate that the centralized \gls{iQPTransformer} achieves \gls{HRLLC} targets beyond the 95th percentile under the \gls{3GPP} channel model with two different mobility models, clearly reflecting the benefits of using a calibrated tail \gls{pdf}. We further evaluate the scalability of the proposed technique, demonstrating that our approach effectively manages a scaling number of \gls{SA} pairs, a capability enabled by the inverted architecture of the \gls{iQPTransformer}. To adapt to resource-constrained scenarios, a U-shaped split-\gls{iQPTransformer} architecture is introduced, enabling distributed interference prediction with minimal degradation compared to the centralized design, as verified through numerical evaluations. Results show that both centralized and split variants maintain scalability and achieve \gls{BLER} targets beyond the 95th percentile, outperforming baseline predictors. This research underscores the potential of a hybrid approach, integrating statistical and advanced machine learning techniques, to develop proactive and scalable interference predictors for next-generation networks. Future work will focus on validating these findings using a real-time setup in dense 6G networks.


\begin{thebibliography}{10}
\providecommand{\url}[1]{#1}
\csname url@samestyle\endcsname
\providecommand{\newblock}{\relax}
\providecommand{\bibinfo}[2]{#2}
\providecommand{\BIBentrySTDinterwordspacing}{\spaceskip=0pt\relax}
\providecommand{\BIBentryALTinterwordstretchfactor}{4}
\providecommand{\BIBentryALTinterwordspacing}{\spaceskip=\fontdimen2\font plus
\BIBentryALTinterwordstretchfactor\fontdimen3\font minus \fontdimen4\font\relax}
\providecommand{\BIBforeignlanguage}[2]{{%
\expandafter\ifx\csname l@#1\endcsname\relax
\typeout{** WARNING: IEEEtran.bst: No hyphenation pattern has been}%
\typeout{** loaded for the language `#1'. Using the pattern for}%
\typeout{** the default language instead.}%
\else
\language=\csname l@#1\endcsname
\fi
#2}}
\providecommand{\BIBdecl}{\relax}
\BIBdecl

\bibitem{hoffmann2023secure}
M.~Hoffmann, G.~Kunzmann, T.~Dudda, R.~Irmer, A.~Jukan, G.~Macher, A.~Ahmad, F.~R. Beenen, A.~Br{\"o}ring, F.~Fellhauer \emph{et~al.}, ``{A Secure and Resilient 6G Architecture Vision of the German Flagship Project 6G-ANNA},'' \emph{IEEE Access}, vol.~11, pp. 102\,643--102\,660, 2023.

\bibitem{recommendation2023framework}
ITU-R, ``{Framework and overall objectives of the future development of IMT for 2030 and beyond},'' \emph{International Telecommunication Union (ITU) Recommendation (ITU-R)}, 2023.

\bibitem{du2022multi}
X.~Du, T.~Wang, Q.~Feng, C.~Ye, T.~Tao, L.~Wang, Y.~Shi, and M.~Chen, ``{Multi-Agent Reinforcement Learning for Dynamic Resource Management in 6G in-X Subnetworks},'' \emph{IEEE Transactions on Wireless Communications}, vol.~22, no.~3, pp. 1900--1914, 2023.

\bibitem{berardinelli2021extreme}
G.~Berardinelli, P.~Baracca, R.~O. Adeogun, S.~R. Khosravirad, F.~Schaich, K.~Upadhya, D.~Li, T.~Tao, H.~Viswanathan, and P.~Mogensen, ``{Extreme Communication in 6G: Vision and Challenges for ‘in-X’ Subnetworks},'' \emph{IEEE Open Journal of the Communications Society}, vol.~2, pp. 2516--2535, 2021.

\bibitem{adeogun2020towards}
R.~Adeogun, G.~Berardinelli, P.~E. Mogensen, I.~Rodriguez, and M.~Razzaghpour, ``{Towards 6G in-X subnetworks with sub-millisecond communication cycles and extreme reliability},'' \emph{IEEE Access}, vol.~8, pp. 110\,172--110\,188, 2020.

\bibitem{schmidt2021interference}
J.~F. Schmidt, U.~Schilcher, M.~K. Atiq, and C.~Bettstetter, ``{Interference prediction in wireless networks: {S}tochastic geometry meets recursive filtering},'' \emph{IEEE Transactions on Vehicular Technology}, vol.~70, no.~3, pp. 2783--2793, 2021.

\bibitem{prameshjournal}
P.~Gautam, C.~Bockelmann, and A.~Dekorsy, ``{Probabilistic Interference Prediction for Dynamic 6G In-X Sub-Networks},'' \emph{IEEE Open Journal of the Communications Society}, vol.~6, pp. 2454--2473, 2025.

\bibitem{clavier2020experimental}
L.~Clavier, T.~Pedersen, I.~Larrad, M.~Lauridsen, and M.~Egan, ``{Experimental evidence for heavy tailed interference in the IoT},'' \emph{IEEE Communications Letters}, vol.~25, no.~3, pp. 692--695, 2020.

\bibitem{padilla2021nonlinear}
C.~Padilla, R.~Hashemi, N.~H. Mahmood, and M.~Latva-Aho, ``{A Nonlinear Autoregressive Neural Network for Interference Prediction and Resource Allocation in URLLC Scenarios},'' in \emph{2021 International Conference on Information and Communication Technology Convergence (ICTC)}, 2021, pp. 184--189.

\bibitem{jayawardhana2023predictive}
C.~Jayawardhana, T.~Sivalingam, N.~H. Mahmood, N.~Rajatheva, and M.~Latva-Aho, ``{Predictive resource allocation for URLLC using empirical mode decomposition},'' in \emph{2023 Joint European Conference on Networks and Communications \& 6G Summit (EuCNC/6G Summit)}, 2023, pp. 174--179.

\bibitem{gunarathne2023decomposition}
S.~Gunarathne, T.~Sivalingam, N.~H. Mahmood \emph{et~al.}, ``{Decomposition Based Interference Management Framework for Local 6G Networks},'' in \emph{2023 IEEE Globecom Workshops}.\hskip 1em plus 0.5em minus 0.4em\relax IEEE, 2023, pp. 1633--1638.

\bibitem{wei2024joint}
L.~Wei, Z.~Liu, C.~Zhang, L.~Zhang, Y.~Wang, and Y.~Huang, ``{Joint Model and Data-Driven Two-Stage Uplink Interference Prediction in URLLC Scenarios},'' in \emph{2024 IEEE Wireless Communications and Networking Conference (WCNC)}.\hskip 1em plus 0.5em minus 0.4em\relax IEEE, 2024, pp. 1--6.

\bibitem{mahmood2020predictive}
N.~H. Mahmood, O.~A. López, H.~Alves, and M.~Latva-Aho, ``{A Predictive Interference Management Algorithm for URLLC in Beyond 5G Networks},'' \emph{IEEE Communications Letters}, vol.~25, no.~3, pp. 995--999, 2021.

\bibitem{brighente2022interference}
A.~Brighente, J.~Mohammadi, P.~Baracca, S.~Mandelli, and S.~Tomasin, ``{Interference prediction for low-complexity link adaptation in beyond 5G ultra-reliable low-latency communications},'' \emph{IEEE Transactions on Wireless Communications}, vol.~21, no.~10, pp. 8403--8415, 2022.

\bibitem{Menholt2022Interference}
L.~Menholt and L.~V. Friis, ``{Mathematical Modelling and Prediction of Interference Power in In-robot Subnetworks},'' Master's thesis, Mathematical Engineering-Aalborg University, 2022.

\bibitem{marzban2023deep}
M.~F. Marzban, W.~Nam, T.~Luo, A.~Kannan, and T.~Yoo, ``Deep learning for probabilistic interference predictions in mmwave networks,'' in \emph{2023 IEEE International Conference on Communications Workshops (ICC Workshops)}.\hskip 1em plus 0.5em minus 0.4em\relax IEEE, 2023, pp. 42--47.

\bibitem{gautamcooperative}
P.~Gautam, M.~Vakilifard, C.~Bockelmann, and A.~Dekorsy, ``{Cooperative Interference Estimation Using LSTM-Based Federated Learning for In-X Subnetworks},'' in \emph{GLOBECOM 2023-2023 IEEE Global Communications Conference}, 2023, pp. 1338--1344.

\bibitem{pramesh2024int}
P.~Gautam, C.~Bockelmann, and A.~Dekorsy, ``{Interference Prediction in Unconnected In-X Mobile 6G Subnetworks Using a Data-Driven Approach},'' in \emph{2024 IEEE International Conference on Communications Workshops (ICC Workshops)}, 2024, pp. 2046--2052.

\bibitem{bennis2018ultrareliable}
M.~Bennis, M.~Debbah, and H.~V. Poor, ``{Ultrareliable and low-latency wireless communication: Tail, risk, and scale},'' \emph{Proceedings of the IEEE}, vol. 106, no.~10, pp. 1834--1853, 2018.

\bibitem{evt_rad_map}
T.~Kallehauge, A.~E. Kalør, P.~Ramírez-Espinosa, C.~Biscio, and P.~Popovski, ``{Prediction of Rare Channel Conditions using Bayesian Statistics and Extreme Value Theory},'' \emph{IEEE Transactions on Communications}, pp. 1--1, 2025.

\bibitem{salehi2025ultra}
F.~Salehi, A.~Mahmood, S.~Coleri, and M.~Gidlund, ``{Ultra-High Reliability by Predictive Interference Management Using Extreme Value Theory},'' \emph{arXiv preprint arXiv:2501.11704}, 2025.

\bibitem{pramesh_extreme}
P.~Gautam, C.~Bockelmann, and A.~Dekorsy, ``{Extreme Value Theory-based Predictive Interference Management for 6G Subnetworks with Transformer},'' in \emph{ICC 2025 - IEEE International Conference on Communications}.\hskip 1em plus 0.5em minus 0.4em\relax IEEE, 2025.

\bibitem{5G-ACIA}
5G-ACIA, ``{A 5G Traffic Model for Industrial Use Cases},'' 5G-ACIA, Germany, White Paper, November 2019, [Online]. Available: {https://5g-acia.org/media/2021/04/WP\_5G\_5G\_Traffic\_Model\_for\_Industrial\_Use \_Cases\_22.10.19.pdf}.

\bibitem{li2023advanced}
D.~Li, S.~R. Khosravirad, T.~Tao, and P.~Baracca, ``{Advanced frequency resource allocation for industrial wireless control in 6G subnetworks},'' in \emph{2023 IEEE Wireless Communications and Networking Conference (WCNC)}, 2023, pp. 1--6.

\bibitem{cuozzo2022enabling}
G.~Cuozzo, S.~Cavallero, F.~Pase, M.~Giordani, J.~Eichinger, C.~Buratti, R.~Verdone, and M.~Zorzi, ``{Enabling URLLC in 5G NR IIoT networks: A full-stack end-to-end analysis},'' in \emph{2022 Joint European Conference on Networks and Communications \& 6G Summit (EuCNC/6G Summit)}.\hskip 1em plus 0.5em minus 0.4em\relax IEEE, 2022, pp. 333--338.

\bibitem{skapoor2018distributedscheduling}
S.~Kapoor, S.~Sreekumar, and S.~R.~B. Pillai, ``{Distributed Scheduling in Multiple Access With Bursty Arrivals Under a Maximum Delay Constraint},'' \emph{IEEE Transactions on Information Theory}, vol.~64, no.~2, pp. 1297--1316, 2018.

\bibitem{3gpp2019service}
3GPP, ``{Service requirements for the 5G system},'' \emph{3rd Generation Partnership Project (3GPP), Technical Specification (TS) 22.261}, 2019.

\bibitem{cavallero2024coexistence}
S.~Cavallero, F.~Saggese, J.~Shiraishi, S.~R. Pandey, C.~Buratti, and P.~Popovski, ``{Coexistence of Pull and Push Communication in Wireless Access for IoT Devices},'' in \emph{2024 IEEE 25th International Workshop on Signal Processing Advances in Wireless Communications (SPAWC)}, 2024, pp. 841--845.

\bibitem{3gpp2018study}
3GPP, ``Study on communication for automation in vertical domains (cav),'' \emph{Technical Report}, 2018.

\bibitem{xiao2006novel}
C.~Xiao, Y.~R. Zheng, and N.~C. Beaulieu, ``{Novel sum-of-sinusoids simulation models for Rayleigh and Rician fading channels},'' \emph{IEEE Transactions on Wireless Communications}, vol.~5, no.~12, pp. 3667--3679, 2006.

\bibitem{3gpp2020study}
3GPP, ``{{5G}: {Study} on channel model for frequencies from 0.5 to 100 {GHz}}~(\textsc{3GPP TR} 38.901 version 16.1.0 release 16),'' 2020.

\bibitem{lu2015effects}
S.~Lu, J.~May, and R.~J. Haines, ``{Effects of correlated shadowing modeling on performance evaluation of wireless sensor networks},'' in \emph{2015 IEEE 82nd Vehicular Technology Conference (VTC2015-Fall)}, 2015, pp. 1--5.

\bibitem{marzban2024interference}
M.~F.~A. Marzban, W.~Nam, T.~Luo, T.~Yoo, and A.~C. Kannan, ``{Interference data collection with beam information for ml-based interference prediction},'' Mar.~14 2024, {US} Patent App. 17/932,191.

\bibitem{polyanskiy2010channel}
Y.~Polyanskiy, H.~V. Poor, and S.~Verd{\'u}, ``{Channel coding rate in the finite blocklength regime},'' \emph{IEEE Transactions on Information Theory}, vol.~56, no.~5, pp. 2307--2359, 2010.

\bibitem{xu2013improving}
X.~Xu, M.~Ni, and R.~Mathar, ``{Improving QoS by predictive channel quality feedback for LTE},'' in \emph{2013 21st International Conference on Software, Telecommunications and Computer Networks-(SoftCOM 2013)}, 2013, pp. 1--5.

\bibitem{herdin2005correlation}
M.~Herdin, N.~Czink, H.~Ozcelik, and E.~Bonek, ``{Correlation matrix distance, a meaningful measure for evaluation of non-stationary MIMO channels},'' in \emph{2005 IEEE 61st Vehicular Technology Conference}, vol.~1.\hskip 1em plus 0.5em minus 0.4em\relax IEEE, 2005, pp. 136--140.

\bibitem{gehring2001empirical}
A.~Gehring, M.~Steinbauer, I.~Gaspard, and M.~Grigat, ``{Empirical channel stationarity in urban environments},'' in \emph{Proceedings of the European Personal Mobile Communications Conference}, 2001.

\bibitem{georgiou2007distances}
T.~T. Georgiou, ``{Distances and Riemannian metrics for spectral density functions},'' \emph{IEEE Transactions on Signal Processing}, vol.~55, no.~8, pp. 3995--4003, 2007.

\bibitem{wang2019probabilistic}
Y.~Wang, D.~Gan, M.~Sun, N.~Zhang, Z.~Lu, and C.~Kang, ``{Probabilistic individual load forecasting using pinball loss guided LSTM},'' \emph{Applied Energy}, vol. 235, pp. 10--20, 2019.

\bibitem{romano2019conformalized}
Y.~Romano, E.~Patterson, and E.~Candes, ``{Conformalized quantile regression},'' \emph{Advances in neural information processing systems}, vol.~32, 2019.

\bibitem{vovk2005algorithmic}
V.~Vovk, A.~Gammerman, and G.~Shafer, \emph{Algorithmic learning in a random world}.\hskip 1em plus 0.5em minus 0.4em\relax Springer, 2005, vol.~29.

\bibitem{xu2023conformal}
C.~Xu and Y.~Xie, ``{Conformal prediction for time series},'' \emph{IEEE Transactions on Pattern Analysis and Machine Intelligence}, 2023.

\bibitem{stankeviciute2021conformal}
K.~Stankeviciute, A.~M~Alaa, and M.~van~der Schaar, ``{Conformal time-series forecasting},'' \emph{Advances in neural information processing systems}, vol.~34, pp. 6216--6228, 2021.

\bibitem{shafer2008tutorial}
G.~Shafer and V.~Vovk, ``{A Tutorial on Conformal Prediction.}'' \emph{Journal of Machine Learning Research}, vol.~9, no.~3, 2008.

\bibitem{kath2021conformal}
C.~Kath and F.~Ziel, ``{Conformal prediction interval estimation and applications to day-ahead and intraday power markets},'' \emph{International Journal of Forecasting}, vol.~37, no.~2, pp. 777--799, 2021.

\bibitem{papadopoulos2002inductive}
H.~Papadopoulos, K.~Proedrou, V.~Vovk, and A.~Gammerman, ``{Inductive confidence machines for regression},'' in \emph{Machine Learning: ECML 2002: 13th European Conference on Machine Learning Helsinki, Finland, August 19--23, 2002 Proceedings 13}.\hskip 1em plus 0.5em minus 0.4em\relax Springer, 2002, pp. 345--356.

\bibitem{jensen2022ensemble}
V.~Jensen, F.~M. Bianchi, and S.~N. Anfinsen, ``{Ensemble Conformalized Quantile Regression for Probabilistic Time Series Forecasting},'' \emph{IEEE Transactions on Neural Networks and Learning Systems}, vol.~35, no.~7, pp. 9014--9025, 2024.

\bibitem{liu2023itransformer}
Y.~Liu, T.~Hu, H.~Zhang, H.~Wu, S.~Wang, L.~Ma, and M.~Long, ``{iTransformer: Inverted Transformers Are Effective for Time Series Forecasting},'' \emph{arXiv preprint arXiv:2310.06625}, 2023.

\bibitem{nie2022time}
Y.~Nie, N.~H.~Nguyen, P.~Sinthong, and J.~Kalagnanam, ``{A Time Series is Worth 64 Words: Long-term Forecasting with Transformers},'' in \emph{International Conference on Learning Representations}, 2023.

\bibitem{vaswani2017attention}
A.~Vaswani, N.~Shazeer, N.~Parmar, J.~Uszkoreit, L.~Jones, A.~N. Gomez, L.~u. Kaiser, and I.~Polosukhin, ``{Attention is All you Need},'' in \emph{Advances in Neural Information Processing Systems}, vol.~30.\hskip 1em plus 0.5em minus 0.4em\relax Curran Associates, Inc., 2017.

\bibitem{kim2021reversible}
T.~Kim, J.~Kim, Y.~Tae, C.~Park, J.-H. Choi, and J.~Choo, ``{Reversible instance normalization for accurate time-series forecasting against distribution shift},'' in \emph{International Conference on Learning Representations}, 2021.

\bibitem{Manaswi2018}
N.~K. Manaswi, \emph{RNN and LSTM}.\hskip 1em plus 0.5em minus 0.4em\relax Berkeley, CA: Apress, 2018.

\bibitem{sak2014long}
H.~Sak, A.~Senior, and F.~Beaufays, ``{Long short-term memory based recurrent neural network architectures for large vocabulary speech recognition},'' \emph{arXiv preprint arXiv:1402.1128}, 2014.

\bibitem{haan2006extreme}
L.~Haan and A.~Ferreira, \emph{{Extreme value theory: an introduction}}.\hskip 1em plus 0.5em minus 0.4em\relax Springer, 2006, vol.~3.

\bibitem{lin2024efficient}
Z.~Lin, G.~Zhu, Y.~Deng, X.~Chen, Y.~Gao, K.~Huang, and Y.~Fang, ``Efficient parallel split learning over resource-constrained wireless edge networks,'' \emph{IEEE Transactions on Mobile Computing}, vol.~23, no.~10, pp. 9224--9239, 2024.

\bibitem{datta2020survey}
L.~Datta, ``A survey on activation functions and their relation with xavier and he normal initialization,'' \emph{arXiv preprint arXiv:2004.06632}, 2020.

\bibitem{kay1993fundamentals}
S.~M. Kay, \emph{Fundamentals of statistical signal processing: estimation theory}.\hskip 1em plus 0.5em minus 0.4em\relax Prentice-Hall, Inc., 1993.

\end{thebibliography}
\end{document}